\documentclass[aps,prl,reprint,showpacs,longbibliography, citeonline,superscriptaddress,]{revtex4-2}
\usepackage{amsmath,amssymb,bm}
\usepackage{graphicx}
\usepackage{xspace}
\usepackage{latexsym}
\usepackage{color}
\usepackage{xcolor}
\usepackage{placeins}

\usepackage[colorlinks=true, citecolor={blue!80!black}, urlcolor={blue!50!black}, linkcolor = {blue!80!black}]{hyperref}

\begin{document}
\title{$LT$ scaling in depleted quantum spin ladders}
\author{S.~Galeski}
\email{Corresponding author: Stanislaw.Galeski@cpfs.mpg.de}
\affiliation{Laboratory for Solid State Physics, ETH Z\"{u}rich, 8093 Z\"{u}rich, Switzerland}
\affiliation{Max Planck Institute for Chemical Physics of Solids, N\"{o}thnitzer Strasse
40,01187 Dresden,Germany}
\author{K.~Yu.~Povarov}
\affiliation{Laboratory for Solid State Physics, ETH Z\"{u}rich, 8093 Z\"{u}rich, Switzerland}
\author{D.~Blosser}
\affiliation{Laboratory for Solid State Physics, ETH Z\"{u}rich, 8093 Z\"{u}rich, Switzerland}
\author{S.~Gvasaliya}
\affiliation{Laboratory for Solid State Physics, ETH Z\"{u}rich, 8093 Z\"{u}rich, Switzerland}
\author{R.~Wawrzynczak}
\affiliation{Max Planck Institute for Chemical Physics of Solids, N\"{o}thnitzer Strasse
40,01187 Dresden,Germany}
\affiliation{Institut Laue-Langevin, 6 rue Jules Horowitz, 38042 Grenoble, France}
\author{J.~Ollivier}
\affiliation{Institut Laue-Langevin, 6 rue Jules Horowitz, 38042 Grenoble, France}
\author{J.~Gooth}
\affiliation{Max Planck Institute for Chemical Physics of Solids, N\"{o}thnitzer Strasse
40,01187 Dresden,Germany}
\author{A.~Zheludev}
\email{zhelud@ethz.ch}
\homepage{https://www.neutron.ethz.ch/}
\affiliation{Laboratory for Solid State Physics, ETH Z\"{u}rich, 8093 Z\"{u}rich, Switzerland}

\begin{abstract}
Using a combination of neutron scattering, calorimetry, Quantum Monte
Carlo (QMC) simulations and analytic results we uncover confinement
effects in depleted, partially magnetized quantum spin ladders. We
show that introducing non-magnetic impurities into magnetized spin
ladders leads to the emergence of a new characteristic length $L$
in the otherwise scale-free Tomonaga--Luttinger liquid (serving as the effective low-energy model). This results
in universal $LT$ scaling of staggered susceptibilities. Comparison
of simulation results with experimental phase diagrams of prototypical
spin ladder compounds DIMPY and BPCB yields excellent agreement.
\end{abstract}
\date{\today}

\maketitle

Understanding the interplay of electron correlation and disorder is
one of the central problems of condensed matter physics. The influence
of disorder becomes crucial in one dimension where the non-trivial
topology does not allow for quasi-particles to avoid even the smallest
defects~\cite{Giamarchi2003}. The impact of such interaction can
be especially dramatic in gapless many-body 1D systems realizing a
quantum critical Tomonaga--Luttinger Liquid (TLL) - a one dimensional
analogue of the celebrated Fermi liquid~\cite{Giamarchi2003,Zaliznyak2005,*Lake_NatMat_2005_ChainScaling,Lake2010,Zheludev_JETP_2020_ScalingReview}.
Similarly to the latter, the TLL is characterized by the analogue
of Fermi velocity $v$ that sets the energy scale (conveniently expressed
in units such as Kelvin), and additional exponent $K$ reflecting
the strength and type of interactions in the system.

\begin{figure}
\includegraphics[width=0.5\textwidth]{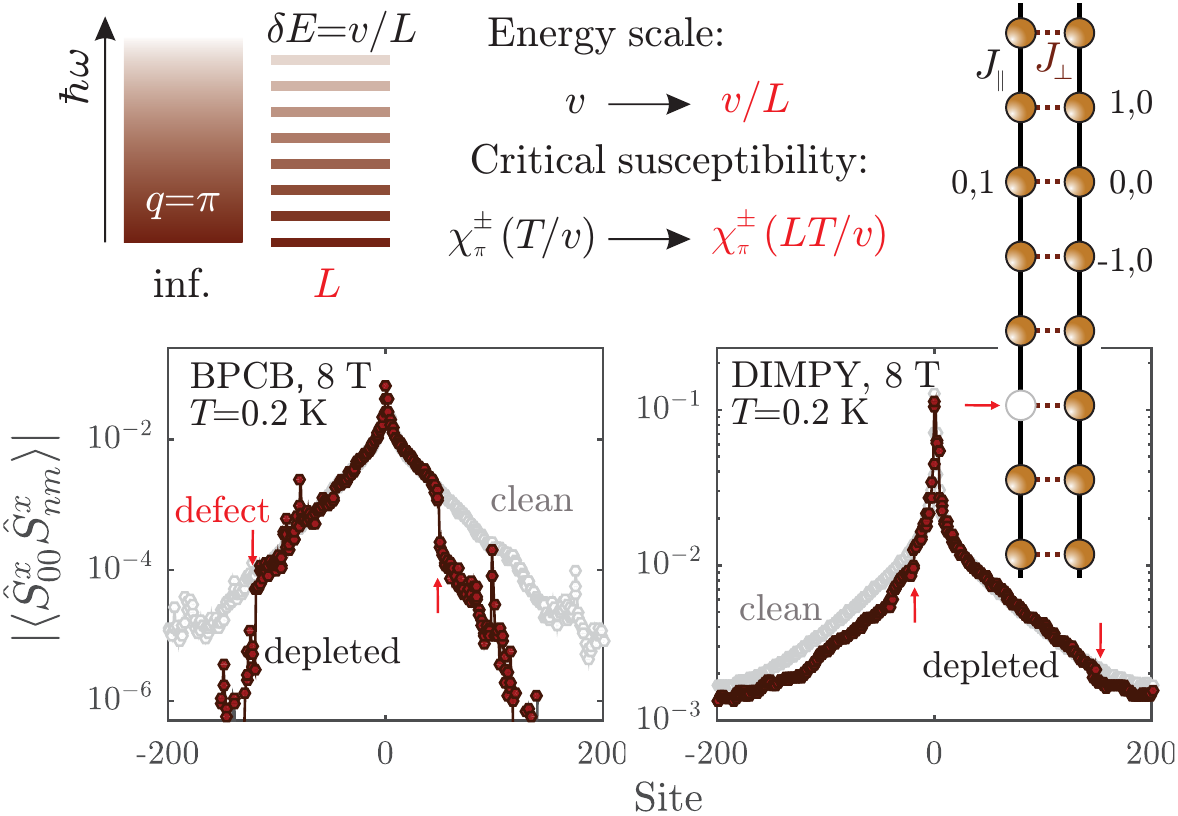} \caption{\label{fig:correlation and continuity}  (a) Origin of $LT$ scaling
in Tomonaga--Luttinger liquids. Finite-size effect results in equidistant
discretization of the energy spectrum at critical wavevector and in
effect to suppression of corresponding susceptibility. (b) A QMC
simulation of the transverse two-spin correlator in a pure and depleted
400-site magnetized DIMPY and BPCB spin ladders. The two randomly placed defects in the
ladder (shown by red arrows) are partially ``transparent'' to correlations. }
\end{figure}

A paradigmatic model for the study of the TLL physics is the simple
$S=1/2$ Heisenberg chain. Recent studies of site disorder in spin
chain materials have shown that introduction of defects into the chain
lattice effectively divides the liquid into finite pieces~\cite{Simutis2013,*Simutis2017,Simutis2016a}.
Such segmentation has a profound effect on the low temperature properties
of the TLL: the pure system, being in a quantum critical state is
characterized by only a single energy scale --- temperature $T$.
Introduction of defects imposes a novel scale: the average distance
between defects $L$ and leads to discretization of the TLL energy
levels with a $v/L$ step [see Fig.~\ref{fig:correlation and continuity}(a)].
This in turn leads to the replacement of the original $T$-universality
by an new $LT$-universality: all dynamic and thermodynamic observables
become functions of the product $LT$ only~\cite{Simutis2016a,Eggert2002a,Sirker2008}. However, for the general anisotropic XXZ case such theoretical studies were performed only in $T=0$ limit so far~\cite{AnnabelleBohrdtKevinJageringSebastianEggert}.

A central question in the study of defects in low dimensional quantum
magnets is whether the $LT$ scaling is a truly universal property
of disordered TLLs, even when defects do not break the liquid's continuity
and for arbitrary TLL exponent $K$. The simplest realistic model
allowing for the study of this problem is the depleted Heisenberg
$S=1/2$ ladder. Unlike the spin chain, it offers a much less restrictive
geometry since removing one spin from the lattice does not break the
ladder continuity, as shown Fig.~\ref{fig:correlation and continuity}.
In addition, different interaction parameters $K$ can be achieved
in ladders with different rung and leg exchange coupling ratios $J_{\perp}/J_{\parallel}$,
with strong leg ladders displaying attractive and strong rung ladders
repulsive interactions~\cite{JeongSchmidiger_PRL_2016_TLSLdichotomy}. This $K$-exponent is also tunable by the magnetic field, and is usually found from numeric simulations for a given system~\cite{HikiharaFurusaki_PRB_2004_LadderKs}. Thus, the depleted spin ladder magnets open up an avenue to explore the “dirty TLL” physics in the most generic, controlled, and tunable way.

\begin{table}
\centering %
\begin{tabular}{lccccccc}
\hline
 & $J_{\perp}$~(K)  & $J_{\parallel}$~(K)  & $J'$~(mK)  & $g_{H\parallel a}$  & $\xi$~(lat. u.)  & $v$~(K)  & $K$ \tabularnewline
\hline
DIMPY  & 9.5  & 16.74  & 75  & 2.13  & 6.3  & 22.04  & 1.23\tabularnewline
BPCB  & 12.96  & 3.6  & 80  & 2.06  & $\sim0.8$  & 3.94  & 0.93\tabularnewline
\hline
\end{tabular}\caption{Magnetic Hamiltonian constants of BPCB~\cite{Bouillot2011} and DIMPY~\cite{Schmidiger2012}
for rung, leg and interladder interaction and the $g$-factor~\cite{Glazkov2015a,Patyal1990} for
the relevant field direction. Also the corresponding correlation length
$\xi$ in the gapped phase, and TLL parameters at $8$~T: Luttinger
velocity $v$ and dimensionless exponent $K$ are given.}
\label{tab:ladders}
\end{table}

In this Letter we experimentally and theoretically study the effect of spin depletion on two
organometallic spin ladder materials: BPCB~\cite{Patyal1990,Thielemann_PRL_2009_BPCBtlsl,Thielemann_PRB_2009_BPCBphasediagram,*Klanjsek_PSS_2010_BPCBnmr,Bouillot2011}
and DIMPY~\cite{Shapiro2007a,Schmidiger2012,Schmidiger_PRL_2013_DimpyLET,Jeong2013,*Povarov2015}.
Both of these materials are well established realizations of $S=1/2$
Heisenberg ladder Hamiltonian and harbour an attractive (DIMPY) and
repulsive (BPCB) TLL when magnetized~\cite{Bouillot2011,Schmidiger2012,JeongSchmidiger_PRL_2016_TLSLdichotomy}.
The most relevant parameters of the materials are given in Table~\ref{tab:ladders}~\cite{SM}. Using
inelastic neutron scattering we experimentally confirm the
applicability of depleted ladder model to Zn$^{2+}$ ($S=0$) - substituted
BPCB. Using a combination of QMC simulations~\cite{Bauer2011}, analytical
calculations, and an extensive low temperature heat capacity survey
we show that the $LT$ scaling of staggered susceptibilities
is a robust, universal property of disordered TLL regardless of the
$K$ parameter. Our findings suggest that such scaling persists even if the continuity of TLL remains preserved, a situation inaccessible in the segmented spin chain settings studied before~\cite{Simutis2016a,Eggert2002a,Sirker2008,AnnabelleBohrdtKevinJageringSebastianEggert}. Thus, $LT$ universality appears to be a generic property of ``dirty'' TLLs.

The samples were grown in house by the thermal gradient {[}(C$_{7}$H$_{10}$N)$_{2}$Cu$_{1-z}$Zn$_{z}$Br$_{4}$
DIMPY{]} and solvent evaporation {[}(C$_{5}$H$_{12}$N)$_{2}$Cu$_{1-z}$Zn$_{z}$Br$_{4}$
BPCB{]} methods, simmilar to the pristine materials~\cite{Patyal1990,Shapiro2007a},
with replacement of the relevant amount $z$ of CuBr$_{2}$ by ZnBr$_{2}$.
Both in specific heat and neutron scattering experiments we have used
fully deuterated chemicals for growth of BPCB crystals. The Zn$^{2+}$
content in the studied samples was verified using Micro X-ray fluorescence
chemical analysis and independently crosschecked through comparison of
low field magnetization measurements with QMC simulations. The DIMPY
samples were already used in the previous study~\cite{Schmidiger2016}.

In a depleted spin ladder, at low magnetic fields the non-magnetic
defects were shown to lead to the emergent clusters of staggered magnetization
(``spin islands''), as typical for a gapped system~\cite{Mikeska1997,*Sigrist1996,*Lavarelo2013,Schmidiger2016}.
The effect of Zn substitution in DIMPY at low fields has been already
thoroughly studied revealing that Zn-doped DIMPY is an excellent realization
of a depleted strong leg spin ladder Hamiltonian~\cite{Schmidiger2016}.
However, despite the structural and chemical similarity between BPCB
and DIMPY, an assumption that BPCB also realizes a depleted spin ladder
Hamiltonian cannot be taken for granted. In order to verify that this
is the case we have studied the emergent Zn-induced ``spin island''
degrees of freedom in the gapped phase of BPCB. Thanks to the short
correlation length in BPCB, these ``islands'' are expected to exhibit
no overlap for low Zn substitution levels and thus behave as uncorrelated
paramagnetic impurities and scatter neutrons elastically. Their presence
could thus be only detected as a magnetic field induced Zeeman resonance.
The localized spin-1/2 degrees of freedom should appear as a non-dispersive
level, very broad in momentum, centered at $\hbar\omega=g\mu_{\mathrm{B}}\mu_{0}H$.
A similar kind of Zeeman resonance has been observed in a depleted
Haldane material Y$_{2}$BaNi$_{1-z}$Mg$_{z}$O$_{5}$~\cite{Kenzelmann_PRL_2003_HaldaneHoles}.

\begin{figure}
\includegraphics[width=0.5\textwidth]{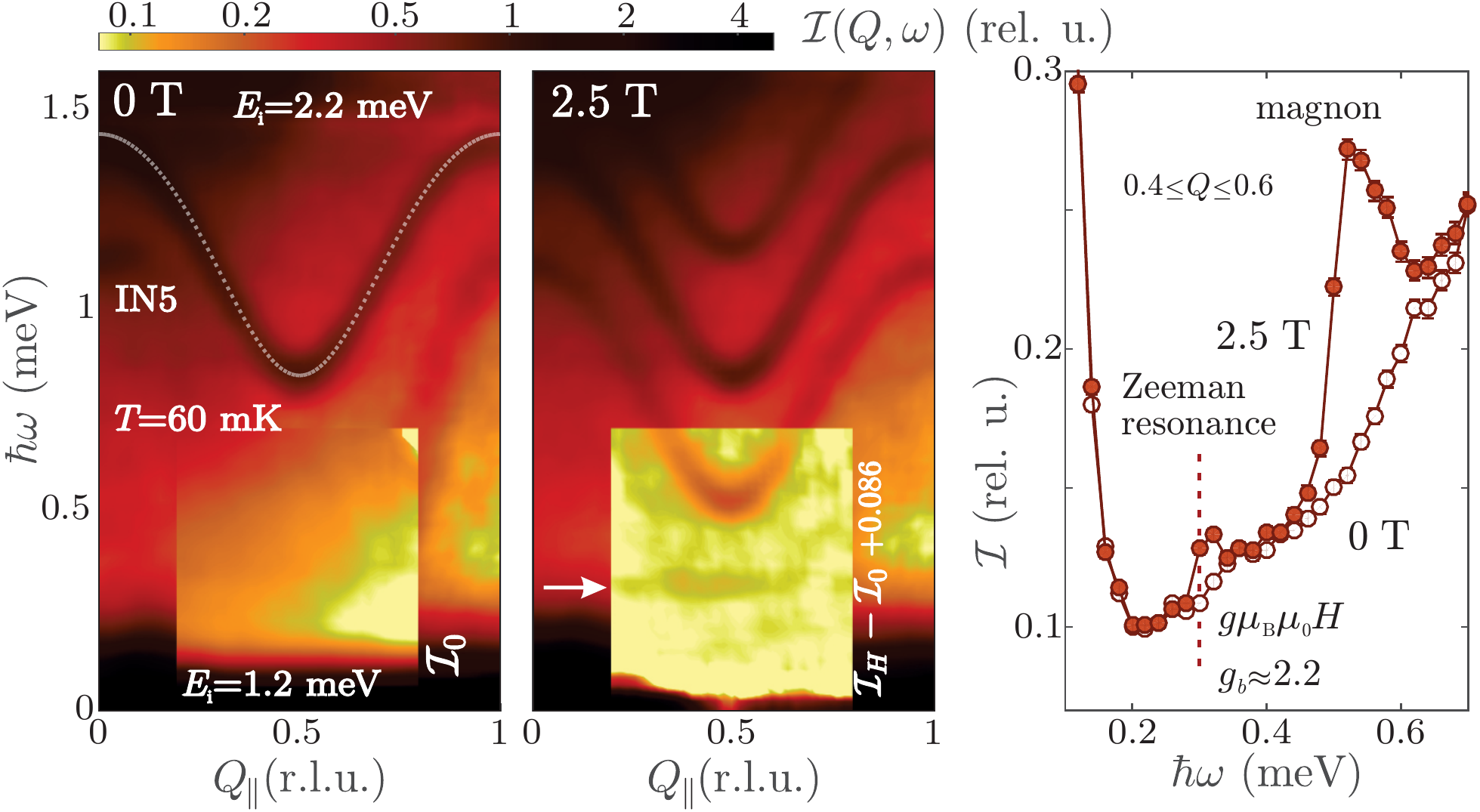} \caption{\label{fig:Magnetic-neutron-scattering} (a,b) Magnetic neutron
scattering intensity as a function of energy and momentum transfer
(along the ladder leg) of depleted BPCB in $0$ and $2.5$ T. The
spectra were taken using an incident neutron energy of $2.2$ (in
the main panels) and $1.2$~meV (in the insets). For $2.5$~T case
inset shows the background-subtracted intensity; the arrow shows the
position of the impurity-related scattering. Dashed line in $0$~T
panel is the reference magnon dispersion in clean BPCB~\cite{BlosserBhartiya_PRB_2019_BPCBanisotropy}. (c) Intensity
profiles $\mathcal{I}(\omega)$ near $Q_{\parallel}=0.5$~r.l.u.
Zeeman resonance at $0.3$~meV is well visible in the magnetized
case. }
\end{figure}

In order to test those predictions we have performed an inelastic
neutron spectroscopy study of 5 coaligned $z=0.02$ BPCB crystals
with total mass $0.9$~g, at the IN5 time-of-flight spectrometer~\cite{OllivierMutka_JPSJ_2011_IN5}
(ILL, Grenoble) with a split-coil vertical magnet. Fig.~\ref{fig:Magnetic-neutron-scattering}(a,b)
shows a comparison between magnetic neutron spectra collected at $T=60$~mK
in and in $0$~T and $2.5$~T with the magnetic field applied along
the crystals $b$-axis. The data were were collected with $2.2$ and
$1.2$~meV incident energy, then treated with \textsc{Horace} software~\cite{Ewings_NucInstr_2016_HORACE}.
Neutron intensity is shown as a function of energy transfer $\hbar\omega$
and momentum transfer along the leg direction $Q_{\parallel}$. The
$0$~T spectrum demonstrates that introduction of non-magnetic defects
does not alter the magnon dispersion~\cite{BlosserBhartiya_PRL_2018_BPCBz2,*BlosserBhartiya_PRB_2019_BPCBanisotropy}
in any obvious way compared with the pristine material. The dispersion itself is well-captured by the strong-rung limit expansion of the spin ladder model~\cite{ReigrotzkiTsunetsuguRice_JPCM_1994_LadderStrongRungExpansion}. There are
no novel low energy features visible over the magnet-related background, justifying the weakly correlated picture of the defects in BPCB, contrasting the case of DIMPY. In contrast at $2.5$~T there appears a clearly distinguishable
non-dispersive level at $0.3$~meV [see Fig.~~\ref{fig:Magnetic-neutron-scattering}(b,c)] --- the expected resonance. The $S=1$ magnon's dispersion remains unchanged apart from splitting into three branches due to Zeeman effect. This is exactly the behavior
one would expect from the depleted spin ladder with the parameters
of BPCB. The availablity of the ``isolated island'' analytic solution in the strong-rung limit with short correlations is very helpful, as the numeric simulations of spectral properties are extremely demanding in the presence of disorder. Thus, we are confident that the chemical substitution in
both materials functions the same way (merely depleting the spin ladder)
and further comparison of critical properties in high magnetic fields
would be meaningful.

In magnetic fields strong enough to close the magnon gap, fate of
the spin impurities is expected to change. Above the critical field
the spin ladder enters the gapless TLL phase~\cite{Giamarchi2003,Giamarchi1999}.
The most distinct difference between TLLs realized in spin chain and
magnetized spin ladders is that ladder lattices offer a much less
restrictive geometry. Thus it can be expected that introduction of
non-magnetic defects could induce a novel length scale into the problem
without sacrificing the TLL continuity. To test if non-magnetic defects
would remain at least partially transparent to spin correlations in
magnetized spin ladders we have performed direct Quantum Monte Carlo
(QMC) simulations of the transverse components of spin-spin correlations$\left\langle \hat{S}_{00}^{x}\hat{S}_{mn}^{x}\right\rangle $
on two ladder lattices with coupling constants resembling that of
our target materials. The calculations have been performed using
the \texttt{dirloop-sse} algorithm of the ALPS package~\cite{Bauer2011,Sandvik_PRB_1999_dirloopQMC,*AletWesselTroyer_PRE_2005_dirloopsseQMC}.
The results of those calculations displayed in Fig.~\ref{fig:correlation and continuity}
suggest that as naively expected, spin correlations can propagate along
the ladder despite the presence of non-magnetic defects for both DIMPY
and BPCB. Comparing the left and right panel of Fig.~\ref{fig:correlation and continuity}(b)
reveal that defects suppress spin correlations much more efficiently
in the case of the repulsive TLL realized in BPCB than in the attractive
case of DIMPY.

In order to verify that introduction of defects that do not lead to
segmentation of the TLL still leads to the presence of $LT$-scaling
in the critical (transverse staggered $\chi_{\pi}^{\pm}$) susceptibility
of magnetized depleted ladders, we have performed another series of
QMC simulations. Although the ALPS \texttt{dirloop-sse} algorithm~\cite{Bauer2011,Sandvik_PRB_1999_dirloopQMC,*AletWesselTroyer_PRE_2005_dirloopsseQMC}
does not give direct access to $\chi_{\pi}^{\pm}$ for magnetized
systems, it does allow for the calcualtion of the static transverse
correlation function $\left\langle S_{00}^{+}S_{nm}^{-}\right\rangle $.
While this quantity is in general not directly related to the susceptibility,
in the case of TLL it can be shown that per mol of spins we have:

\begin{equation}
\chi_{\pi}^{\pm}(L,T)=\frac{N_{A}}{k_{B}}\frac{\left(g\mu_{B}\right)^{2}}{LT\mathcal{R}\left(K\right)}\sum_{n,m,k,l}\left(-1\right)^{n-m+k-l}\left\langle S_{nk}^{+}S_{ml}^{-}\right\rangle, \label{eq:qmc2chi}
\end{equation}

with $\mathcal{R}(K)$ being a prefactor of the order of $1$, that
can be analytically derived~\cite{SM}.

\begin{figure}
\includegraphics[width=0.5\textwidth]{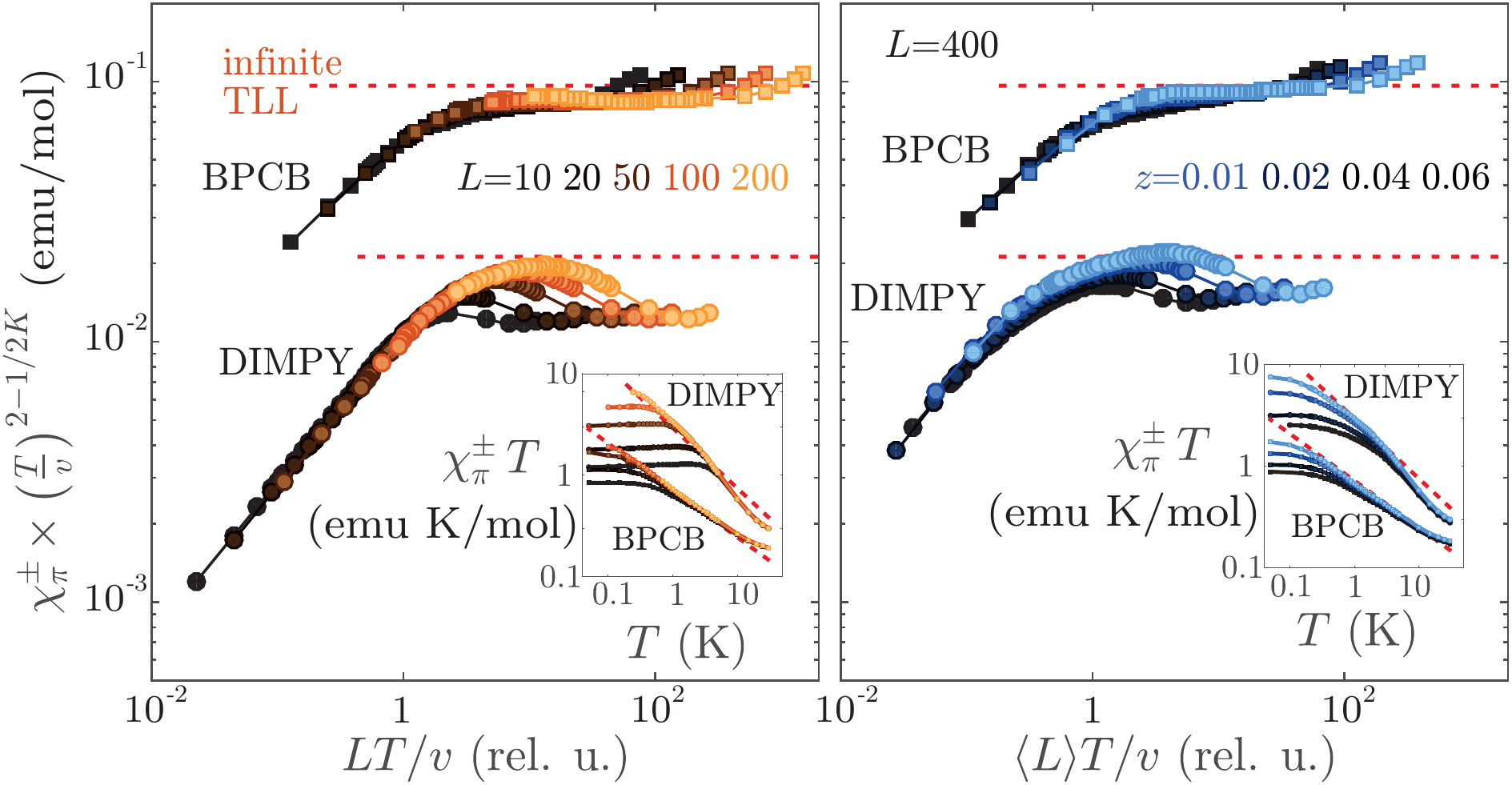} \caption{\label{fig:Staggered QMC} Comparison of $LT/v$ scaling properties
of staggered susceptibilities in finite spin ladder segments and depleted
spin ladders. Susceptibility is extracted from QMC data according
to (\ref{eq:qmc2chi}); simulations are done at $8$~T field for
coupling constants of DIMPY and BPCB. (a) Scaling of finite spin
ladder segments with open boundary conditions.  (b)  Scaling of $400$
rung diluted spin ladder lattices, here $L=\frac{1-z}{2z-z^{2}}$
is the average inter-defect distance. In both panels the dashed lines
show the scaling law expected for the infinite system. Insets shows
the unscaled data. }
\end{figure}

Using the QMC results and Eq.~(\ref{eq:qmc2chi}) we have computed
the staggered transverse susceptibility for a series of spin ladder
lattices of $10$, $20$, $50$, $100$ and $200$ rungs with coupling
constants resembling those of DIMPY and BPCB, in $8$~T field, and
with the boundaries being open. The resulting susceptibilities are
plotted in the inset of Fig.~\ref{fig:Staggered QMC}(a).
However, as we can see, the simple $\chi_{\pi}^{\pm}(L,T)\times T$
product does not show apparent scaling as it would in the open boundaries
Heisenberg spin chain case with $K=1/2$~\cite{Eggert2002a}. The
temperature-related power law requires a modification here, and we
argue that the generalized scaling relation should be~\cite{SM}:

\begin{equation}
\chi_{\pi}^{\pm}(L,T)\propto\left(\frac{T}{v}\right)^{\frac{1}{2K}-2}F_{K}\left(\frac{LT}{v}\right),\label{eq:ansatz}
\end{equation}

with $F_{K}(x)$ being a $K$-parameterized single argument function.
Once the data is replotted in the respective coordinates [Fig.~\ref{fig:Staggered QMC}(a)], the presence of scaling~(\ref{eq:ansatz})
in the low-$T$ regime becomes apparent. This is consistent with the power law $T^{1/2K-2}$ for the thermodynamic limit~\cite{Giamarchi2003}. At higher temperatures the behavior is non-TLL in general and the scaling breaks down.

To test the hypothesis that introducing random spin depletion also
introduces a new length scale leading to $LT$ scaling we have performed
a second set of QMC simulations. Assuming the same coupling parameters
and applied magnetic fields we have computed the susceptibilities
for both ladder systems with 400 rungs and 1, 2, 4 and 6\% spin depletion
and averaged over 20 different random impurity distributions. The
results of these calculations are displayed in Fig.~\ref{fig:Staggered QMC}(b).
Strikingly despite some differences in overall shape of the curves
compared to the finite length pieces, the susceptibilities in ``depleted
ladders'' closely follow the derived scaling law.

These theoretical results would be best confirmed through a direct
comparison with measurements of transverse staggered susceptibilities
of several samples with different level of dilution. Although such
measurements could be in principle carried out using inelastic neutron
scattering, in practice it would be a formidable task. Another avenue
for testing our QMC results can be provided through testing the ability
of our target systems to achieve 3D ordering. One of the most spectacular
ways spin depletion influences low temperature properties of quasi-1D
magnets is through altering their ordering capabilities: in ultra
pure site diluted SrCuO$_{2}$ samples the presence of less than 1\%
of impurities suppresses N\'{e}el ordering beyond experimental detection~\cite{Simutis2016a}.
The mean field condition for magnetic ordering ties the ordering temperature,
critical susceptibility and the strength of residual 3D couplings
$J'$~\cite{Schulz_PRL1996_coupledchains,Giamarchi1999,Yasuda2005}:

\begin{equation}
\chi_{\pi}^{\pm}\left(T_{N}\right)=\frac{N_{A}}{k_{B}}\frac{(g\mu_{B})^{2}}{J'}.\label{eq:mean field-1}
\end{equation}

This explains the strong suppresion of $T_{N}$ in depleted spin chains:
addition of impurities effectively cuts them, truncating the spin-spin
correlations and leading to depression of the relevant susceptibility.
Eq.~(\ref{eq:mean field-1}) defines the critical susceptibility
magnitude that triggers long range ordering. Thus to confirm the validity
of our QMC simulations we have calculated the full phase diagram of
both pure and depleted DIMPY and BPCB for a direct comparison with
experimental phase diagrams obtained from specific heat measurements.
We would like to stress that the simulations and measurements are
performed away from the critical fields, and so possible Bose-glass
related phenomena~\cite{TrinhHaas_PRB_2013_BGladders} are not relevant.

\begin{figure}
\includegraphics[width=0.5\textwidth]{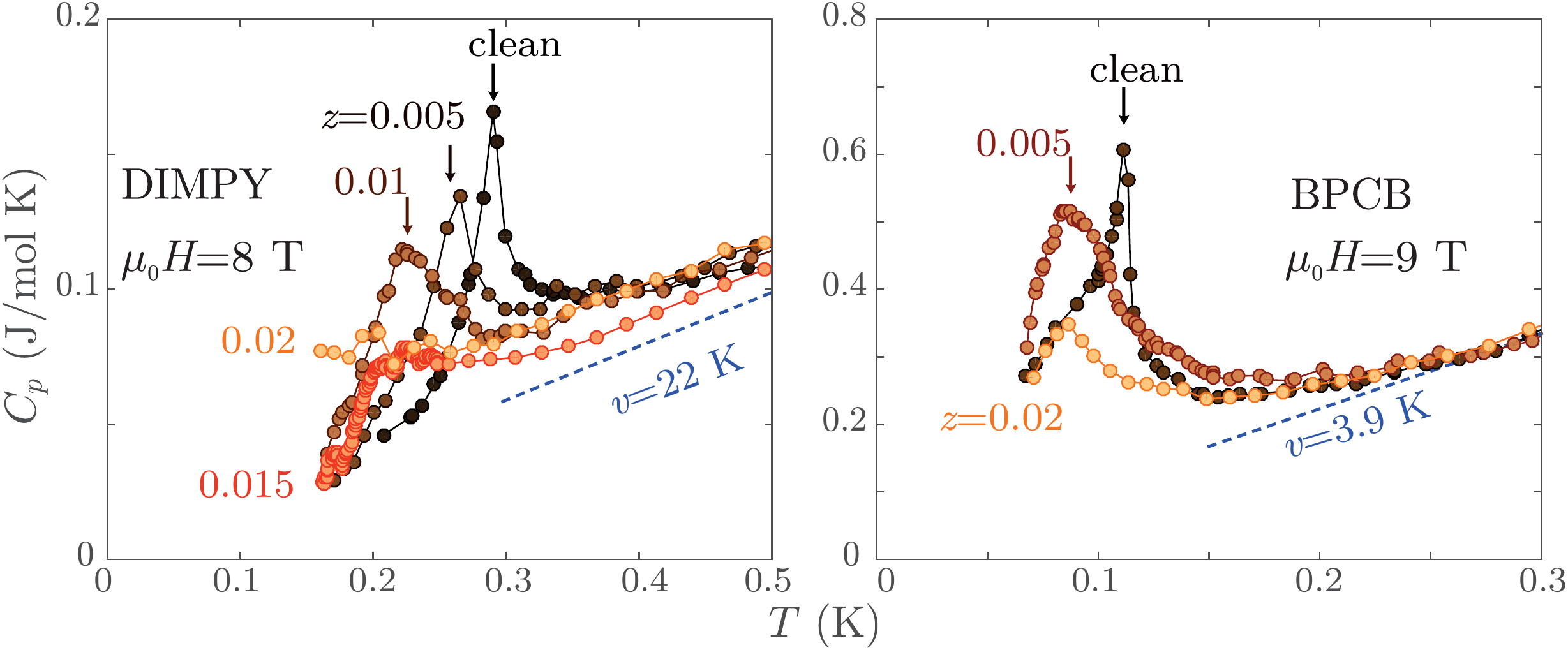} \caption{\label{fig:Heat Capacity} Comparison of the spin dilution effect
on the specific heat anomaly of DIMPY and BPCB measured at $8$ and
$9$~T, respectively. Dashed lines show the expected TLL linear specific
heat. Arrows mark the transition temperature. }
\end{figure}

The magnetic phase diagrams for all samples were determined through
locating the maxima of the specific heat anomalies attributed to 3D
ordering via fitting Lorentzian-like peak functions (details are shown in~\cite{SM}). The specific
heat measurements were performed in the Quantum Design PPMS dilution
refrigerator equipped in a standard specific heat option and a $9$~T
superconducting magnet. Representative datasets after subtracting
the nuclear spin contribution for samples with different Zn substitution
level of both DIMPY and BPCB are shown in Fig.~\ref{fig:Heat Capacity}.
Inspection of both datasets reveals that spin depletion apart form
quickly suppressing magnetic order additionally leads to broadening
of the specific heat anomalies. It was shown that such broadening
can be an effect of weak random fields arising due to chemical
disorder~\cite{WulfMuhlbauer_PRB_2011_SulDisordered}. Indeed it
can be expected that spin depletion of a single ladder not only decreases
its staggered transverse susceptibility leading to the depression
of $T_{N}$ but also exerts an effective random field on the neighbouring
ladders. Interestingly apart from altering the shape and location
of the specific heat anomaly spin depletion seems to have no influence
on the high temperature specific heat. This simple observation suggests
that although adding non-magnetic impurities changes the span of correlation
functions it does not influence the spinon velocity $v$ as $C_{v}^{\mathrm{TLL}}(T)=R\frac{\pi}{6v}T$.
Moreover, previous studies of equal-time correlations scaling in depleted
chains~\cite{Simutis2013,*Simutis2017} suggest that defects do not
alter the exponent $K$ either, in agreement with our QMC analysis.

\begin{figure}
\includegraphics[width=0.5\textwidth]{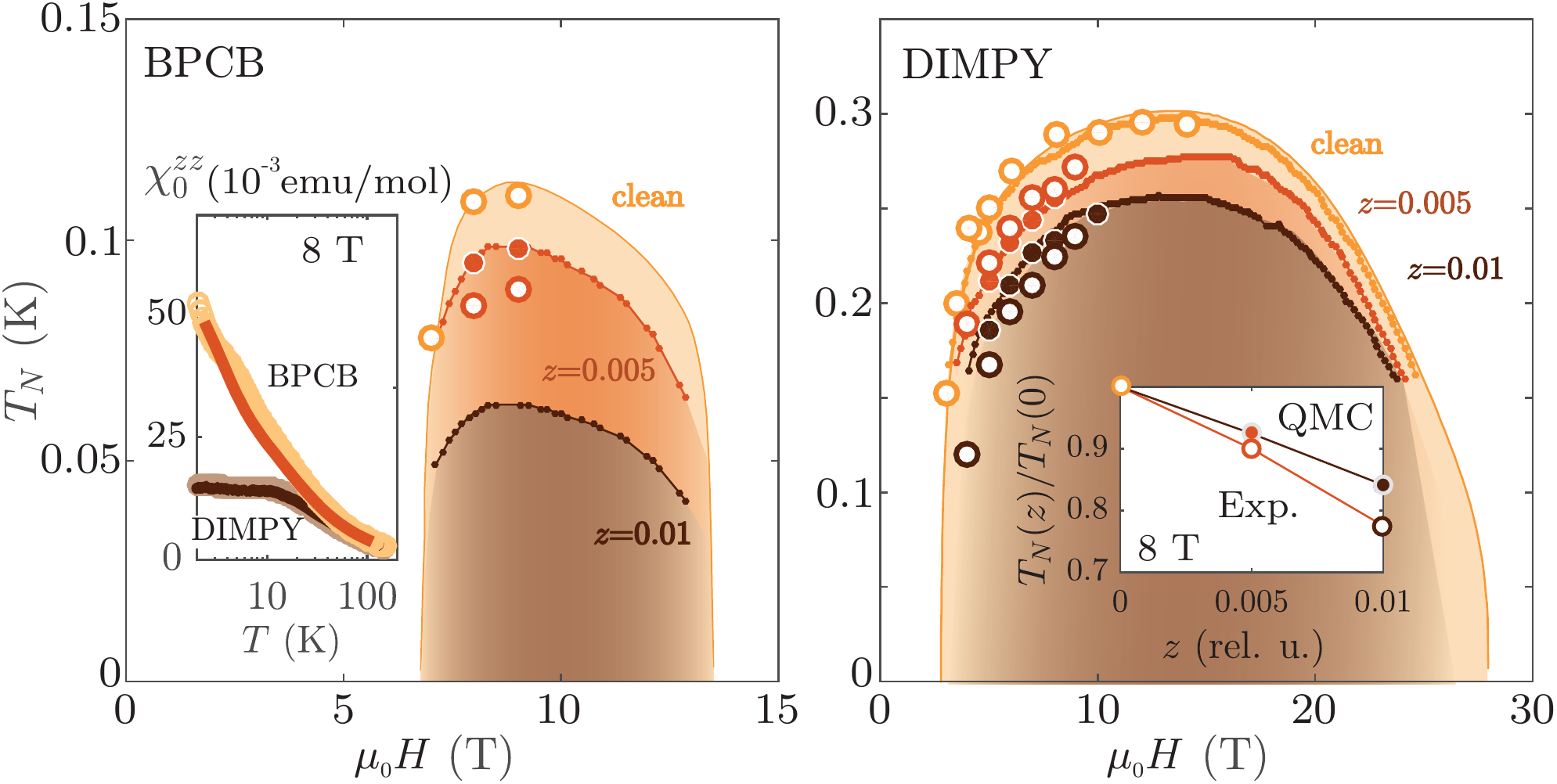} \caption{\label{fig:phase diagrams} Comparison of QMC simulation results with
experimental phase diagrams obtained from specific heat: large solid
circles mark points in the $H-T$ plane where the calculation was
performed using $100$ random impurity configurations; small circles
are the rescaled results with $20$ configurations. The large open
circles represent results of heat capacity measurements. Thin line
for the clean case is the DMRG-based mean field result. (a) The inset shows a comparison of simulated and measured uniform longitudinal
magnetic susceptibility per spin of pure DIMPY and BPCB at $8$~T
(lines --- QMC results, symbols are the measured values). (b) The inset shows a comparison of experimental and simulated
values of $T_{N}$ at $8$~T.   }
\end{figure}

To verify that our QMC procedure is reliable in computing physical
properties of the relevant materials we have calculated the longitudinal
uniform susceptibility which is a directly observable quantity. For
comparison with simulations we have measured the susceptibility of
both materials using the PPMS AC-susceptibility option at $8$~T.
As shown in the inset of Fig.~\ref{fig:phase diagrams}(a) the agreement
of the QMC result with experiment is excellent confirming the validity
of our approach.

The first crucial step in simulating phase diagrams of depleted spin
ladders is the determination of the inter ladder coupling constant
$J'$. To do this we have calculated the transverse staggered susceptibility
for pure 400 rung systems with periodic boundry conditions and then
fitted Eq.~(\ref{eq:mean field-1}) to the experimental specic heat data
using $J'$ as a fitting parameter, following the procedure of \cite{Schmidiger2012}.
As a result we have established the interladder coupling to be 106mK
and 110~mK for DIMPY and BPCB respectively, values slightly higher
than previously obtained through a combined DMRG and field theory
approach. Although estimating $T_{N}$ for the case of depleted ladders
form susceptibility calculations with similar level of averaging to
those used for the scaling analysis qualitatively reproduce the experimental
phase diagram, the predicted phase boundaries showed to be off by
over 20\% compared to the experimental results. In order to correct
for the effect of under-averaging we have thus calculated the expected
$T_{N}$ for several fields averaging over 100 random impurity configurations.
This allowed us to obtain predictions for $T_{N}$ with excellent
agreement with empirical data.

The final comparison of experimental and calculated phase diagrams
is shown in the main panels of Fig.~\ref{fig:phase diagrams}. For
both materials the QMC result seems to accurately reproduce the phase
diagrams for both pure and diluted samples. The excellent predictive
power of our QMC calculation confirms that our simulation method can
be used to accurately calculate $\chi_{\pi}^{\pm}(T)$ for a range
of magnetic fields. This in turn reassures the validity of the $LT$
scaling analysis in depleted spin ladders. Finally, we would like
to note that in case of attractive TLL in DIMPY the ordered phase
is much more resistant to defects compared to the case of repulsive
TLL in BPCB. This seems to be the experimental consequence of spontaneous
defect ``healing'' in an attractive TLL~\cite{Kane1992}, also
in line with the direct calculation of (small) correlation suppression
in DIMPY case shown in Fig.~\ref{fig:correlation and continuity}(b).

Our results combined with Ref.~\cite{Schmidiger2016} paint a full
picture of the interplay of spin depletion and magnetic fields in
spin ladders. At low magnetic fields $H<H_{c1}$ defects introduce
packets of localized staggered magnetization that can give rise to
a new emergent magnetic system if located densely enough to interact
with one another. On the other hand at high fields $H>H_{c1}$ defects
truncate the spin-spin correlation introducing a new length scale
and introducing the new $LT$ universality even when the TLL is not
fully segmented. Those results suggests that the quantum critical
state is in general susceptible to adjusting to externally imposed
length scales such as disorder. We have verified that the $LT$ scaling generally emerges in the presence of impurities for any degree of continuity and any Luttinger exponent $K$. Our analysis is expected
to be directly transferable to other TLL systems such as nanowires,
or the quantum Hall edge states. Another interesting possible application are the 1D disordered quantum magnets with charge degrees of freedom, such as Sr$_{14}$Cu$_{24}$O$_{41}$~\cite{BlumbergLittlewood_Science_ladderCDW,*SahlingRemenyi_NPhys_2015_Entanglement} or BaFe$_2$S$_3$~\cite{TakahashiSugimoto_NatMat_2015_BaFe2S3pressed,*HirataMaki_PRB_2015_BaFe2S3doped}.
It remains to be verified if our results
will hold in these more complex situations.
\begin{acknowledgments}
This work was supported by the Swiss National Science Foundation,
Division II. Part of the numerical simulations were performed on the
ETH Z\"{u}rich Euler cluster. We also acknowledge Mr. Jonathan Noky for
his assistance in running the calculations on the computing cluster
of Max Planck Institute for Chemical Physics of Solids, Dresden. We
would like to thank Dr. D. Schmidiger (ETH Z\"{u}rich) for his involvement
at the early stage of the project, Dr. R. Chitra (ETH Z\"{u}rich) and Dr. Tobias Meng (TU Dresden) for
insightful discussions, and Dr. F. Alet (CNRS, Universit\'{e} Paul Sabatier
Toulouse) for helpful advice on QMC techniques.
\end{acknowledgments}

\bibliography{DIMPYBiblio_v30}

\end{document}


\title{Supplemental Material for\\ ``$LT$ scaling in depleted quantum spin ladders''}
\author{S.~Galeski}
\email{Corresponding author: Stanislaw.Galeski@cpfs.mpg.de}
\affiliation{Laboratory for Solid State Physics, ETH Z\"{u}rich, 8093 Z\"{u}rich, Switzerland}
\affiliation{Max Planck Institute for Chemical Physics of Solids, N\"{o}thnitzer Strasse
40,01187 Dresden,Germany}
\author{K.~Yu.~Povarov}
\affiliation{Laboratory for Solid State Physics, ETH Z\"{u}rich, 8093 Z\"{u}rich, Switzerland}
\author{D.~Blosser}
\affiliation{Laboratory for Solid State Physics, ETH Z\"{u}rich, 8093 Z\"{u}rich, Switzerland}
\author{S.~Gvasaliya}
\affiliation{Laboratory for Solid State Physics, ETH Z\"{u}rich, 8093 Z\"{u}rich, Switzerland}
\author{R.~Wawrzynczak}
\affiliation{Max Planck Institute for Chemical Physics of Solids, N\"{o}thnitzer Strasse
40,01187 Dresden,Germany}
\affiliation{Institut Laue-Langevin, 6 rue Jules Horowitz, 38042 Grenoble, France}
\author{J.~Ollivier}
\affiliation{Institut Laue-Langevin, 6 rue Jules Horowitz, 38042 Grenoble, France}
\author{J.~Gooth}
\affiliation{Max Planck Institute for Chemical Physics of Solids, N\"{o}thnitzer Strasse
40,01187 Dresden,Germany}
\author{A.~Zheludev}
\email{zhelud@ethz.ch}
\homepage{https://www.neutron.ethz.ch/}
\affiliation{Laboratory for Solid State Physics, ETH Z\"{u}rich, 8093 Z\"{u}rich, Switzerland}

\begin{abstract}
In this Supplemental Material we discuss the possibility of extracting the transverse staggered susceptibility of the magnetized ladder $\chi^{\pm}_\pi$ from the QMC data and the attempt to analytically describe such using the concept of universality and $LT$-scaling for the finite segments. A comparison between the numeric, analytical, and experimental results for the depleted ladders is also given. We also discuss the inelastic neutron scattering experiment and the treatment of specific heat data in some more details.
\end{abstract}
\date{\today}

\maketitle
\tableofcontents

\section{Phase diagrams of depleted ladders}

The extensive survey of low temperature specific heat of depleted
DIMPY and BPCB allows to compile effective phase diagrams relating
the transition temperature, spin depletion and magnetic field. Due
to the disorder induced broadening of the transitions and the small
data density precise determination of the transition temperature through
direct inspection of the specific heat curves seems to be rather difficult.
In order to overcome this difficulty we have extracted the critical
temperature through fitting an empirical ``peak function'' composed
of two Lorentzian-like functions. One of them is multiplied by $T$ in order
to account for the asymmetric appearance of the specific heat anomalies:

\begin{equation}
P(T)=\frac{a_{1}}{(T-T_{N})^{2}+\gamma_{1}}+\frac{a_{2}T}{(T-T_{N})^{2}+\gamma_{2}}\label{eq:Fitting double lorenzian}
\end{equation}

\begin{figure}
\includegraphics[width=0.5\textwidth]{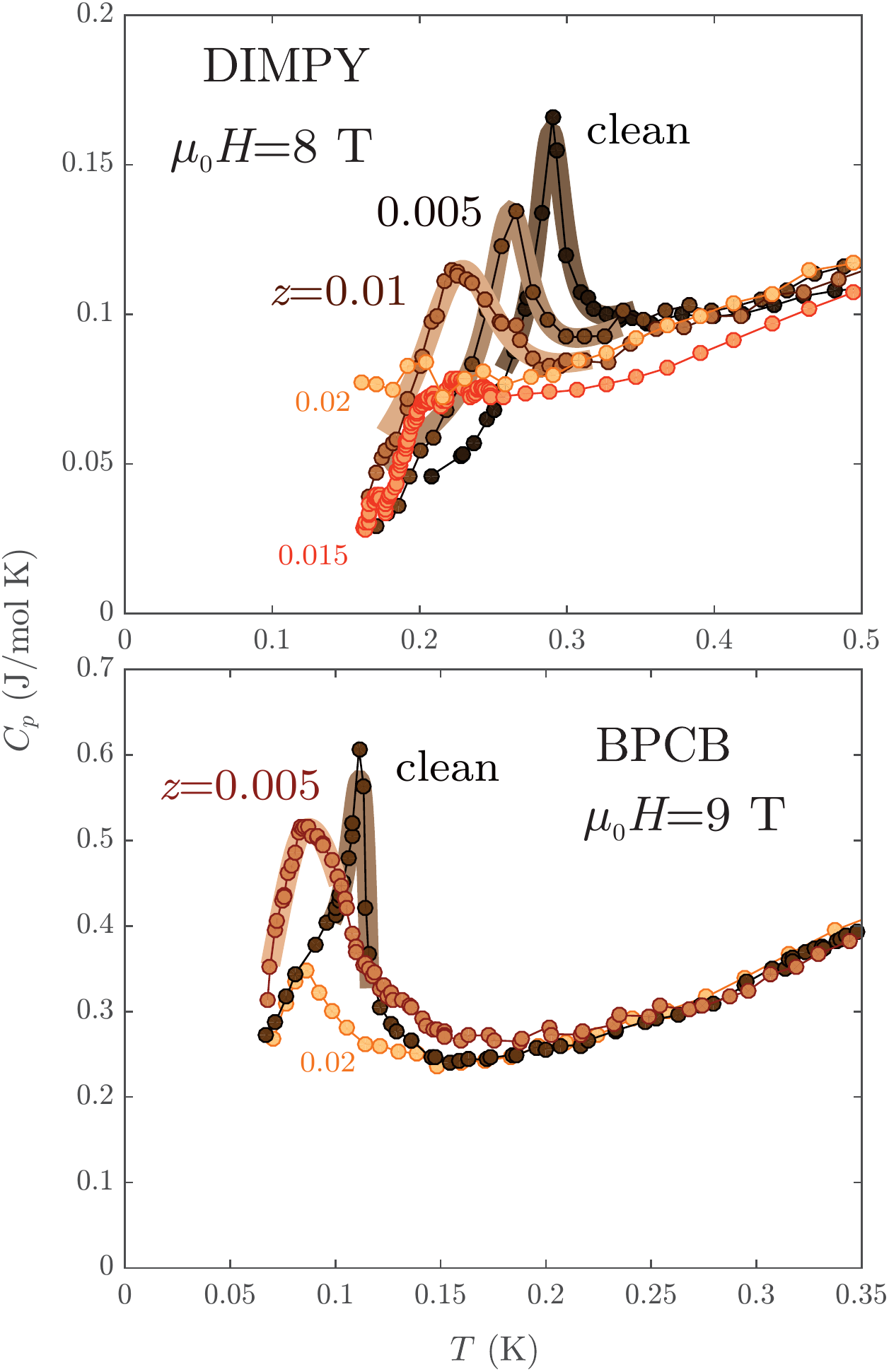}
\caption{\label{fig:Equation--was}Eq.~(\ref{eq:Fitting double lorenzian}) [solid lines]
was used to fit the specific heat data [points] in the vicinity of $T_{N}$
in order to systematically extract the critical temperature from experimental
data.}
\end{figure}

Example results of the fitting procedure is shown in Fig.~\ref{fig:Equation--was}.
The fits obtained in the case of DIMPY allow to extract the value
of $T_{N}$ quite precisely. In the case of BPCB extracting $T_{N}$ proved
to be more difficult due to the complex shape of the specific heat
anomalies. Although a satisfactory overlap of the experimental data
with the model was obtained in the close vicinity of $T_{N}$,
in the case of the 0.5\% Zn sample this result should be treated with
some prudence due to the possible influence of the anomaly
tentatively attributed to nuclear specific heat. Despite this complication
a careful subtraction of the specific heat curve of the 2\% Zn substituted
d-BPCB allows to estimate with high confidence that the presence of
the additional ordering anomaly should not affect the extracted value
of $T_{c}$ of the Cu ions by more than 5-6\%.

\section{Inelastic neutron scattering experiment}

The inelastic neutron scattering experiment was performed on the time-of-flight
spectrometer IN5 at Institute Laue--Lagnevin, Grenoble, France. The
sample consisted of five coaligned deuterated BPCB crystals with 2\%
Zn subsitution, with total mass of the assembly being nearly $0.9$~g.
Crystals were mounted on the aluminum sampleholder with $\mathbf{b}$
axis being vertical. The sampleholder was installed in a $^{3}$He-$^{4}$He
dilution cryostat equipped with a split-coil vertical $2.5$~T magnet.
At the fields of both $0$ and $2.5$~T the experiment was performed
in in high resolution ($E_{\mathrm{i}}=1.2$~meV, $\delta E ~ 20 \mu$eV FWHM)  and low resolution ($E_{\mathrm{i}}=2.2$~meV, $\delta E ~ 45 \mu$eV) modes.
In both modes the sample was rotated by $150^{\circ}$ with
a $2^{\circ}$ step. Neutron exposure time was nearly 35 minutes at
every frame.

The neutron intensities measured at a given momentum transfer $\mathbf{Q}$
and energy transfer $\hbar\omega$ are directly proportional to $\sum_{\alpha,\beta}(1-Q^{\alpha}Q^{\beta}/|\mathbf{Q}|^{2})\mathcal{S}^{\alpha\beta}(\mathbf{Q},\omega)$.
The measured dynamic structure factors $\mathcal{S}^{\alpha\beta}(\mathbf{Q},\omega)$
are the Fourier transforms of the corresponding correlation functions:

\begin{equation}
\mathcal{S}^{\alpha\beta}(\mathbf{Q},\omega)=\frac{1}{2\pi\hbar}\int\limits _{-\infty}^{+\infty}\left\langle \hat{S}^{\alpha}(\mathbf{Q},0)\hat{S}^{\beta}(\mathbf{-Q},t)\right\rangle e^{-i\omega t}dt\label{EQ:sqw}
\end{equation}

The full dataset at each particular field and $E_{\text{i}}$ is thus
the intensity $\mathcal{I}(\mathbf{Q},\omega)$, with $\mathbf{Q}=h\mathbf{a}^{\ast}+k\mathbf{b}^{\ast}+l\mathbf{c}^{\ast}$,
being within the $(h,0,l)$ plane predominantly. The collected data
was projected onto the ladder direction $\mathbf{a}$. Thus, in Fig.~2
of the main text the intensity is shown as the function of momentum
transfer along the ladder $Q_{\parallel}=(\mathbf{Q}\cdot\mathbf{a})/2\pi$
only.

\section{Quantum Monte Carlo simulations}

The Quantum Monte--Carlo simulations have been performed by means
of Stochastic Series Expansion algortihm (\verb"dirloop-sse") of
the ALPS library~\cite{Sandvik_PRB_1999_dirloopQMC,Bauer2011}. The
calculations were carried out on Euler cluster (ETH Z\"{u}rich) and on
the computing cluster of Max Planck Institute for Chemical Physics
of Solids, Dresden. A thermalization of $10^{6}$ was used, with $5\cdot10^{4}$
cycles per point. For the $400$-rung depleted ladder model an averaging
between $20$ disorder configurations was done.

In Table~\ref{TAB:ladders} all the relevant parameters for both
DIMPY and BPCB are summarized. Only $J_{\perp}$, $J_{\parallel}$,
and the magnetic field strength are the input parameters of the simulations.

\begin{table}
\centering %
\begin{tabular}{rlclc}
\hline\hline
 &  & DIMPY  &  & BPCB \tabularnewline
\hline
$J_{\parallel}$~(K)  &  & 16.74~\cite{Schmidiger2012}  &  & 3.6~\cite{Bouillot2011} \tabularnewline
$J_{\perp}$~(K)  &  & 9.5~\cite{Schmidiger2012}  &  & 12.96~\cite{Bouillot2011} \tabularnewline
$J'$~(mK)  &  & 75~\cite{Schmidiger2012}  &  & 80~\cite{Bouillot2011} \tabularnewline
$g_{a}$  &  & 2.13~\cite{Glazkov2015a}  &  & 2.06~\cite{Patyal1990} \tabularnewline
$\xi$ ($a$ units)  &  & 6.3~\cite{Schmidiger2016}  &  & 0.8~\cite{Lavarelo2013} \tabularnewline
\hline
Simulated $H$~(T)  &  & 8  &  & 8 \tabularnewline
\hline
$K$  &  & 1.23~\cite{Schmidiger2012}  &  & 0.93~\cite{Bouillot2011} \tabularnewline
$v$~(K)  &  & 22.04~\cite{Schmidiger2012}  &  & 3.94~\cite{Bouillot2011} \tabularnewline
$A_{x}$  &  & 0.1891~\cite{Schmidiger2012}  &  & 0.1581~\cite{Bouillot2011} \tabularnewline
$\dfrac{1}{2K}-2$  &  & -1.59  &  & -1.46 \tabularnewline
$\mathcal{R}$  &  & 1.17  &  & 1.35 \tabularnewline
\hline\hline
\end{tabular}\caption{Basic Hamiltonian parameters of DIMPY and BPCB along with the relevant
parameters of the Tomonaga--Luttinger liquid description in the $8$~T
magnetic field.}
\label{TAB:ladders}
\end{table}

\section{Extracting the transverse staggered susceptibility}

In this section we will show that as long as the TLL-type universality
holds in the system, it is possible to reliably estimate the static
susceptibility from the equal-time correlations alone, without explicitly
calculating the time-dependent ones.

\subsection{Fluctuation-Dissipation Theorem and Kramers--Kronig relations}

The \verb"dirloop-sse" algorithm from the ALPS software provides
access to the \emph{equal time spin-spin correlation function}:

\begin{align}
\mathcal{C}^{xx}(nk,ml) & =\frac{3}{S(S+1)}\langle\langle\hat{S}_{nk}^{x}\hat{S}_{ml}^{x}\rangle\rangle\\
 & =4\langle\langle\hat{S}_{nk}^{x}(0)\hat{S}_{ml}^{x}(0)\rangle\rangle=2\langle\langle\hat{S}_{nk}^{+}(0)\hat{S}_{ml}^{-}(0)\rangle\rangle\nonumber
\end{align}

Here the double angular brackets $\langle\langle\hat{A}\rangle\rangle$
stand for the statistical average $\mathrm{Tr}\left(\hat{\rho}\hat{A}\right)$,
where $\hat{\rho}$ is the density matrix of the system.

For mapping the ladder onto the TLL we place the pseudospin objects
$\hat{\mathbf{s}}_{n}$ on each rung:

\begin{equation}
\hat{S}_{nk}^{z}=\frac{1}{4}+\frac{1}{2}\hat{s}_{n}^{z}\mathrm{~and~}\hat{S}_{nk}^{\pm}=\frac{(-1)^{k}}{\sqrt{2}}\hat{s}_{n}^{\pm}\label{EQ:pseudospin}
\end{equation}

For the pseudospins the following \emph{equal time spin structure
factor} would be relevant:

\begin{equation}
\mathcal{S}_{\pi}^{\pm}=\sum\limits _{n,m}\langle\langle\hat{s}_{n}^{+}\hat{s}_{m}^{-}\rangle\rangle(-1)^{n-m}
\end{equation}

As $\langle\langle\hat{s}_{n}^{+}\hat{s}_{m}^{-}\rangle\rangle=\langle\langle(\hat{S}_{n1}^{x}-\hat{S}_{n2}^{x})(\hat{S}_{m1}^{x}-\hat{S}_{m2}^{x})\rangle\rangle$,
we can relate this structure factor to the results of the numeric
calculation:

\begin{align}
\mathcal{S}_{\pi}^{\pm}=\sum_{n,m,k,l}(-1)^{n-m+l-k}\langle\langle\hat{S}_{nk}^{x}\hat{S}_{ml}^{x}\rangle\rangle\label{EQ:PseusoALPS}\\
=\frac{1}{4}\sum_{n,m,k,l}(-1)^{n-m+l-k}\mathcal{C}^{xx}(nk,ml)
\end{align}

An alternative way to define the equal time structure factor is to
frequency-integrate the complete \emph{spin dynamic structure factor}:

\begin{equation}
\mathcal{S}_{\pi}^{\pm}=\int\limits _{-\infty}^{+\infty}\mathcal{S}_{\pi}^{\pm}(\omega)d(\hbar\omega),
\end{equation}

that is:
\begin{widetext}
\begin{equation}
\mathcal{S}_{\pi}^{\pm}(\omega)=\int\limits _{-\infty}^{+\infty}\langle\langle\hat{s}^{+}(\pi,0)\hat{s}^{-}(-\pi,t)\rangle\rangle e^{-i\omega t}dt=\int\limits _{-\infty}^{+\infty}\sum\limits _{n,m}\langle\langle\hat{s}_{n}^{+}(0)\hat{s}_{m}^{-}(t)\rangle\rangle(-1)^{n-m}e^{-i\omega t}dt
\end{equation}
\end{widetext}

The spin dynamic structure factor is related to the \emph{dissipative
part of the susceptibility} via the \emph{fluctuation-dissipation
theorem}:

\begin{equation}
\chi_{\pi}^{\pm''}(\omega)=\frac{\pi(g\mu_{B})^{2}}{L}\left[1-e^{-\hbar\omega/k_{B}T}\right]\mathcal{S}_{\pi}^{\pm}(\omega)
\end{equation}

This susceptibility is defined per rung, hence the ladder length $L$
in the denominator.

Finally, the \emph{reactive part of the susceptibility} can be obtained
with the help of \emph{Kramers--Kronig relations}:

\begin{equation}
\chi_{\pi}^{\pm'}(\omega)=\frac{1}{\pi}\mathcal{P}\int\limits _{-\infty}^{+\infty}\frac{\chi_{\pi}^{\pm''}(\Omega)}{\Omega-\omega}d\Omega
\end{equation}
\\
 The static $\omega=0$ limit of the susceptibility is thus:
\begin{widetext}
\begin{equation}
\chi_{\pi}^{\pm'}=\frac{1}{\pi}\mathcal{P}\int\limits _{-\infty}^{+\infty}\frac{\chi_{\pi}^{\pm''}(\Omega)}{\Omega}d\Omega=\frac{(g\mu_{B})^{2}}{L}\mathcal{P}\int\limits _{-\infty}^{+\infty}\frac{\left[1-e^{-\hbar\Omega/k_{B}T}\right]}{\Omega}\mathcal{S}_{\pi}^{\pm}(\Omega)d\Omega\label{EQ:KKintegral}
\end{equation}
\end{widetext}

This is the rigorous formula recovering the \emph{Van Vleck part}
of the isothermal susceptibility. In the present case this is the
only term constituting the susceptibility of interest $\chi_{\pi}^{\pm}(T)=2 \chi_{\pi}^{\pm'}$
(factor of 2 is from the fact that there are 2 spins per rung).

\subsection{Application to the Tomonaga--Luttinger Liquid}

We would like to start with the discussion of the infinite size system
(the clean case with $L\rightarrow\infty$). Let us have a look at
the Luttinger liquid dynamic structure factor at $q=\pi$ and at a
finite temperature. It has a remarkable \emph{universal} form:

\begin{equation}
\mathcal{S}_{\pi}^{\pm}(\omega)=\frac{A_{x}L}{\pi}(T/v)^{1/2K-2}\Phi\left(\frac{\hbar\omega}{k_{B}T}\right)\label{eq:LLsqw}
\end{equation}

Here $A_{x}$ is some non-universal prefactor, and the scaling function
that also depends on the Luttinger exponent $K$ is:

\begin{equation}
\Phi\left(x\right)=\frac{1}{1-e^{-x}}\mathrm{Im}\left[\left(\frac{\Gamma(1/8K-ix/4\pi)\Gamma(1-1/4K)}{\Gamma(1-1/8K-ix/4\pi)}\right)^{2}\right]
\end{equation}

Another quantity of interest is the equal time structure factor, that
is:
\begin{widetext}
\begin{equation}
\mathcal{S}_{\pi}^{\pm}=\frac{A_{x}L}{\pi}\int\limits _{-\infty}^{+\infty}(T/v)^{1/2K-2}\Phi\left(\frac{\hbar\omega}{k_{B}T}\right)d(\hbar\omega)=\frac{A_{x}L}{\pi}(T/v)^{1/2K-1}\int\limits _{-\infty}^{+\infty}\Phi(x)dx
\end{equation}

What we have to compare is $\mathcal{S}_{\pi}^{\pm}$ and the value
of Kramers--Kronig integral~(\ref{EQ:KKintegral}):
\begin{equation}
L^{-1}\int\limits _{-\infty}^{+\infty}\frac{A_{x}L}{\pi}\dfrac{1-e^{-\hbar\Omega/k_{B}T}}{\Omega}\mathcal{S}_{\pi}^{\pm}(\Omega)d\Omega=\frac{A_{x}}{\pi}(T/v)^{1/2K-2}\int\limits _{-\infty}^{+\infty}\dfrac{1-e^{-\hbar\Omega/k_{B}T}}{\hbar\Omega/k_{B}T}\Phi\left(\frac{\hbar\Omega}{k_{B}T}\right)d(\hbar\Omega/k_{B}T)
\end{equation}
\end{widetext}

Thus, there exists a temperature-independent ratio:

\begin{equation}
\mathcal{R}(K)=\frac{2k_{B}^{-1}(g\mu_{B})^{2}\mathcal{S}_{\pi}^{\pm}}{LT\chi_{\pi}^{\pm}}=\frac{\int\limits _{-\infty}^{+\infty}\Phi(x)dx}{\int\limits _{-\infty}^{+\infty}\frac{1-e^{-x}}{x}\Phi(x)dx}\label{eq:Ratio}
\end{equation}

For a given $K$ this is just a number! Hence one can establish the
exact correspondence between the staggered transverse susceptibility
and the equal-time structure factor. The resulting way to relate the
staggered transverse susceptibility \emph{per spin} and the outcome of QMC calculation
is:
\begin{widetext}
\begin{equation}
\chi_{\pi}^{\pm}=\frac{k_{B}^{-1}(g\mu_{B})^{2}}{2LT\times\mathcal{R}(K)}\sum_{n,m,k,l}(-1)^{n-m+l-k}\mathcal{C}^{xx}(nk,ml),\label{eq:Master}
\end{equation}
\end{widetext}

\begin{figure*}
\centering \includegraphics[width=1\textwidth]{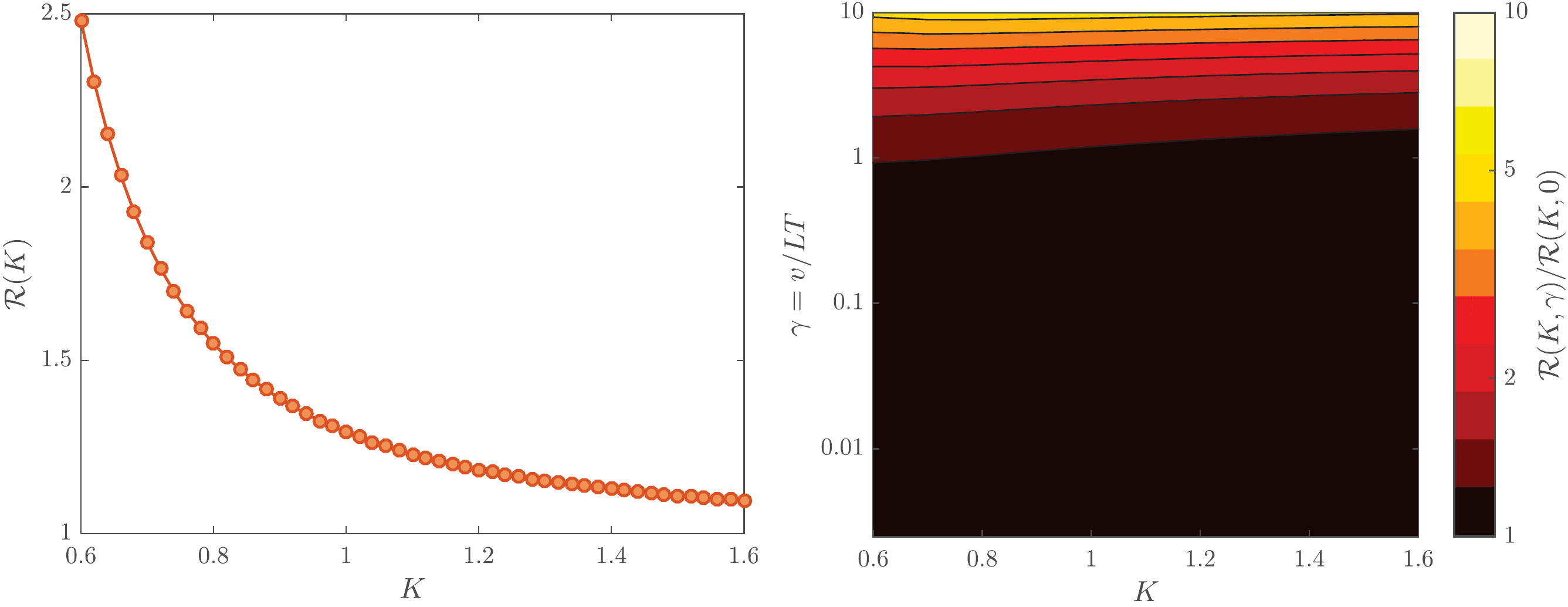} \caption{Left: The ratio~(\ref{eq:Ratio}) for the infinite length case. Right:
The relative change due to the finite-size effects~(\ref{eq:RatioFinite}).}
\label{fig:RK}
\end{figure*}

with the Luttinger exponent $K(H)$ being the calibrated function
of the static longitudinal magnetic field for each particular incarnation
(DIMPY or BPCB) of the ladder model. As one can see from the Figure
\ref{fig:RK}, the value of $\mathcal{R}$ is around $1.5$ for the
$K$'s of interest.

\subsection{Finite-size applicability}

Now we can turn to the discussion of the depleted systems. The corresponding
modification of the structure factor~(\ref{eq:LLsqw}) is:

\begin{equation}
\mathcal{S}_{xx}(\pi,\omega)=A_{x}(T/v)^{1/2K-2}\Phi\left(\frac{\hbar\omega}{k_{B}T}\right)\mathcal{F}\left(\frac{\Delta}{\hbar\omega}\right)
\end{equation}

The \emph{envelope function} describing the missing due to depletion
spectral weight at low energies (the pseudogap) can be approximated
as:

\begin{equation}
\mathcal{F}(y)=\frac{y^{2}}{\sinh(y)^{2}},~\Delta\simeq k_{B}v/\tilde{L}
\end{equation}

Such form of envelope function was used, for instance, in the discussion
of dynamics of the depleted spin chain~\cite{Simutis2013}. Remarkably,
it is temperature independent, and it accounts simply for the discreetness
of the spectrum in the finite-sized segments. Here $v$ is the Luttinger
velocity (setting the overall energy scale of a problem) and the parameter
$\tilde{L}$ can be understood as the typical, or average, segment
length. The ratio $\mathcal{R}$, corresponding to the modified dynamic
structure factor can be straightforwardly evaluated, and it will depend
on the ``magic'' finite-size scaling parameter $\gamma=\frac{v}{\tilde{L}T}$.
This is due to the fact that $\mathcal{F}\left(\frac{v}{\omega\tilde{L}}\right)=\mathcal{F}\left(\frac{\gamma}{\omega/T}\right)$.
Thus, we need to evaluate:

\begin{equation}
\mathcal{R}(K,\gamma)=\frac{\int\limits _{-\infty}^{+\infty}\Phi(x)\mathcal{F}\left(\frac{\gamma}{x}\right)dx}{\int\limits _{-\infty}^{+\infty}\frac{1-e^{-x}}{x}\Phi(x)\mathcal{F}\left(\frac{\gamma}{x}\right)dx}\label{eq:RatioFinite}
\end{equation}

The result is shown in Fig.~\ref{fig:RK}. The conclusion is, finite-size
effects can be neglected as long as $v/\tilde{L}\lesssim T$, that
is mostly the case in our calculations. The ``clean'' prefactor
$\mathcal{R}(K)$ can be used also for the depleted ladders, and Eq.~(\ref{eq:Master})
holds.


\section{Finite-size scaling of transverse staggered susceptibility}

In this section we propose the \emph{ansatz} to derive the transverse
staggered susceptibility of a magnetized TLL. For the sake of brevity
and clearness we omit the factors $\mu_{0}$, $g\mu_{B}$, and $k_{B}$
here, to recover them at the final stages of our calculation.

\subsection{Eggert-Affleck-Horton recipe}

\subsubsection{Basic ideas}

We would like to start with the reminder for the isotropic $S=1/2$
chain at zero field as formulated by Eggert, Affleck and Horton in
Ref.~\cite{Eggert2002a}. We attempt to calculate the ordering temperature
in the mean-field ensemble of chain segments. The distribution of
the segment lengths corresponds to randomly placed chain breaks with
$z$ being the impurity concentration.

As the first step we calculate the single-segment staggered susceptibility,
assuming it to be long enough for ``Luttinger Liquid in a box''
approach to be applicable. The staggered susceptibility in the isotropic
case is:

\begin{equation}
\chi_{\pi}^{\pm}(L,T)=2\chi_{\pi}^{zz}(L,T)=\frac{2}{L}\int\limits _{0}^{1/T}d\tau\iint\limits _{0}^{L}dxdyG(x,y,\tau)\label{EQ:chi_definition}
\end{equation}

It is defined via the Green function:
\begin{widetext}
\begin{equation}
G(x,y,\tau)=A_{z}\frac{\pi}{4L}\frac{\partial_{x}\theta_{1}(0,e^{-\frac{\pi\gamma}{2}})}{\sqrt{\theta_{1}(\frac{\pi x}{L},\exp(-\frac{\pi\gamma}{2}))\theta_{1}(\frac{\pi y}{L},\exp(-\frac{\pi\gamma}{2}))}}\left[G^{+}(x,y,\tau)-G^{-}(x,y,\tau)\right]\label{EQ:EAH_Green}
\end{equation}

where

\begin{equation}
G^{+}(x,y,\tau)=\frac{\theta_{2,3}(\pi\frac{x-y}{L},e^{-\pi\gamma})}{\theta_{2,3}(0,e^{-\pi\gamma})}\sqrt{\frac{\theta_{1}(\frac{\pi}{2}\frac{x+y+iv\tau}{L},e^{-\frac{\pi\gamma}{2}})\theta_{1}(\frac{\pi}{2}\frac{x+y-iv\tau}{L},e^{-\frac{\pi\gamma}{2}})}{\theta_{1}(\frac{\pi}{2}\frac{x-y+iv\tau}{L},e^{-\frac{\pi\gamma}{2}})\theta_{1}(\frac{\pi}{2}\frac{x-y-iv\tau}{L},e^{-\frac{\pi\gamma}{2}})}},\label{EQ:SQRTthingP}
\end{equation}

and

\begin{equation}
G^{-}(x,y,\tau)=\frac{\theta_{2,3}(\pi\frac{x+y}{L},e^{-\pi\gamma})}{\theta_{2,3}(0,e^{-\pi\gamma})}\sqrt{\frac{\theta_{1}(\frac{\pi}{2}\frac{x-y+iv\tau}{L},e^{-\frac{\pi\gamma}{2}})\theta_{1}(\frac{\pi}{2}\frac{x-y-iv\tau}{L},e^{-\frac{\pi\gamma}{2}})}{\theta_{1}(\frac{\pi}{2}\frac{x+y+iv\tau}{L},e^{-\frac{\pi\gamma}{2}})\theta_{1}(\frac{\pi}{2}\frac{x+y-iv\tau}{L},e^{-\frac{\pi\gamma}{2}})}}.\label{EQ:SQRTthingM}
\end{equation}
\end{widetext}

The important scaling parameter $\gamma$ is defined as $\gamma=\frac{v}{LT}$
(with $v=\pi J/2$ in the isotropic chain). The Dedekind theta functions
$\theta_{n}(u,q)$ should be taken as $\theta_{2}$ for odd, and $\theta_{3}$
for even chains. The amplitude prefactor $A_{z}$ is non-universal
(we also omit its weak temperature dependence due to logarithmic corrections).

The remaining ingredient is the partial derivative $\partial_{x}\theta_{1}(0,e^{-\frac{\pi\gamma}{2}})\equiv\frac{\partial\theta_{1}(u,q)}{\partial u}|_{u=0}$.

So-defined segment susceptibility is then used to get the mean staggered
susceptibility value by averaging over the length distribution:

\begin{equation}
\left\langle \chi_{\pi}^{\pm}(T,z)\right\rangle =\sum\limits _{L}\chi_{\pi}^{\pm}(L,T)z^{2}L(1-z)^{L}.\label{EQ:chi_average}
\end{equation}

Then within the mean-field approximation the Neel temperature is given
by:

\begin{equation}
\left\langle \chi_{\pi}^{\pm}(T_{N},z)\right\rangle =\frac{1}{J'}\label{EQ:TN_master}
\end{equation}

\subsubsection{Dimensionless}

Now we can reformulate the above to make the $LT/v$ scaling apparent.
We make the integration variables dimensionless: $\tilde{x}=x/L$,
$\tilde{y}=y/L$ and $\tilde{\tau}=\tau T$. Hence $v\tau/L=\gamma\tilde{\tau}$,
with $\gamma=v/LT$ being our dimensionless parameter of interest.
Then the susceptibility is:

\begin{equation}
\chi_{\pi}^{\pm}(L,T)=2\frac{A_{z}}{T}F_{1/2}^{e,o}(\gamma)\label{EQ:chi_definitionRED}
\end{equation}

where

\[
F_{1/2}^{e,o}(\gamma)=\frac{\pi}{4}\iiint\limits _{0}^{1}\mathcal{G}_{1/2,\gamma}(\widetilde{x},\widetilde{y},\widetilde{\tau})d\widetilde{x}d\widetilde{y}d\widetilde{\tau}
\]
and the function $\mathcal{G}_{1/2,\gamma}$ being:
\begin{widetext}
\begin{equation}
\mathcal{G}_{1/2,\gamma}(\tilde{x},\tilde{y},\tilde{\tau})=\frac{\partial_{x}\theta_{1}(0,e^{-\frac{\pi\gamma}{2}})}{\sqrt{\theta_{1}(\pi\tilde{x},e^{-\frac{\pi\gamma}{2}})\theta_{1}(\pi\tilde{y},e^{-\frac{\pi\gamma}{2}})}}\left[\mathcal{G}^{+}(\tilde{x},\tilde{y},\tilde{\tau})-\mathcal{G}^{-}(\tilde{x},\tilde{y},\tilde{\tau})\right]\label{EQ:EAH_Green_reduced}
\end{equation}

\begin{equation}
\mathcal{G}^{+}(\tilde{x},\tilde{y},\tilde{\tau})=\frac{\theta_{2,3}(\pi(\tilde{x}-\tilde{y}),e^{-\pi\gamma})}{\theta_{2,3}(0,e^{-\pi\gamma})}\sqrt{\frac{\theta_{1}(\frac{\pi(\tilde{x}+\tilde{y}+i\gamma\tilde{\tau})}{2},e^{-\frac{\pi\gamma}{2}})\theta_{1}(\frac{\pi(\tilde{x}+\tilde{y}-i\gamma\tilde{\tau})}{2},e^{-\frac{\pi\gamma}{2}})}{\theta_{1}(\frac{\pi(\tilde{x}-\tilde{y}+i\gamma\tilde{\tau})}{2},e^{-\frac{\pi\gamma}{2}})\theta_{1}(\frac{\pi(\tilde{x}-\tilde{y}-i\gamma\tilde{\tau})}{2},e^{-\frac{\pi\gamma}{2}})}}\label{EQ:SQRTthingPtilde}
\end{equation}

\begin{equation}
\mathcal{G}^{-}(\tilde{x},\tilde{y},\tilde{\tau})=\frac{\theta_{2,3}(\pi(\tilde{x}+\tilde{y}),e^{-\pi\gamma})}{\theta_{2,3}(0,e^{-\pi\gamma})}\sqrt{\frac{\theta_{1}(\frac{\pi(\tilde{x}-\tilde{y}+i\gamma\tilde{\tau})}{2},e^{-\frac{\pi\gamma}{2}})\theta_{1}(\frac{\pi(\tilde{x}-\tilde{y}-i\gamma\tilde{\tau})}{2},e^{-\frac{\pi\gamma}{2}})}{\theta_{1}(\frac{\pi(\tilde{x}+\tilde{y}+i\gamma\tilde{\tau})}{2},e^{-\frac{\pi\gamma}{2}})\theta_{1}(\frac{\pi(\tilde{x}+\tilde{y}-i\gamma\tilde{\tau})}{2},e^{-\frac{\pi\gamma}{2}})}}\label{EQ:SQRTthingMtilde}
\end{equation}
\end{widetext}

Here $e,o$ are standing for the cases of even or odd length (and
thus $\theta_{3}$ or $\theta_{2}$ functions in~(\ref{EQ:SQRTthingPtilde},\ref{EQ:SQRTthingMtilde})).

Having these function tabulated is the key to the calculation.

\subsection{Generalization to the anisotropic case}

\subsubsection{Our assumptions}

Transverse and longitudinal staggered susceptibilities for anisotropic
Luttinger liquid in the magnetic field at $\omega=0$ are known to
be~\cite{Giamarchi2003}:
\begin{widetext}
\begin{equation}
\chi^{zz}(q,T)=\frac{A_{z}}{4}\left[\Phi_{K}(q-\pi(1-2m),T)+\Phi_{K}(q-\pi(1+2m),T)\right]+\frac{1}{4\pi}\Xi_{K}(q),\label{EQ:LL_zz}
\end{equation}

\begin{equation}
\chi^{\pm}(q,T)=A_{x}\Phi_{1/4K}(q-\pi,T).\label{EQ:LL_xx}
\end{equation}
\end{widetext}

Here $m$ is the magnetization in relative units, $q$ is the momentum,
$K$ is the Luttinger parameter. The exact representation of $\Phi_{K}$
and $\Xi_{K}$ functions is of no importance for the discussion below.

A key observation can be made from the generalized susceptibility
definitions~(\ref{EQ:LL_zz},\ref{EQ:LL_xx}). For $m=0$ and $q=\pi$
one has the longitudinal staggered susceptibility $\chi^{zz}(T)=\frac{A_{z}}{2}\Phi_{K}(0,T)$.
At the same time, for arbitrary magnetization $m$ and $q=\pi$ the
transverse staggered susceptibility has a very similar form $\chi^{\pm}(T)=A_{x}\Phi_{1/4K}(0,T)$.
Another term $\Xi_{K}(q)$ is irrelevant close to $q=\pi$. From this
follows our \textbf{assumption 1}: \emph{the staggered transverse
susceptibility of a magnetized TLL with open boundary conditions $\chi_{\pi}^{\pm}=A_{x}f(L,T,\frac{1}{4K})$,
where $f(L,T,K)=2\chi_{\pi}^{zz}/A_{z}$ defines the staggered longitudinal
susceptibility of zero-field TLL with the same Luttinger exponent
$K$ and velocity $v$}. Thus, in a case we have a formula for the
finite-size zero field longitudinal susceptibility at hand, it can
be easily converted to the formula for transverse susceptibility of
a magnetized system.

In principle, our assumption only states the simplest possible form
of the transverse susceptibility, that is compatible with the condition
$\chi_{\pi}^{xx}=\chi_{\pi}^{zz}$ for the finite-size isotropic system.
However, more complicated scenarios are imaginable, for instance $f(L,T,K)=2\chi_{\pi}^{zz}/A_{z}$
and $\chi_{\pi}^{\pm}/A_{x}=f(L\zeta(K),T,\frac{1}{4K})$ with $\zeta(1/2)=1$.
In principle, such renormalization of length might be the reason for
the ``horizontal'' mismatch between the numeric and analytical results
in Fig.~\ref{fig:QMCscaled} of the present Supplement.

A calculation for the anisotropic zero-field case does exist. An approximate
Green function for the longitudinal staggered susceptibility of the
XXZ chain can be found in the recent paper~\cite{AnnabelleBohrdtKevinJageringSebastianEggert}.
This function is written for the limit $\gamma\rightarrow\infty$
--- finite chains and small temperatures (i.e. difference between
the even and odd case is always going to be dramatic). The analogues
of functions (\ref{EQ:EAH_Green},\ref{EQ:SQRTthingP},\ref{EQ:SQRTthingM})
are:
\begin{widetext}
\begin{equation}
G(x,y,\tau)=\frac{A_{z}}{2}\left(\frac{4L^{2}}{\pi^{2}}\sin\frac{\pi x}{L}\sin\frac{\pi y}{L}\right)^{-K}\left[G^{+}(x,y,\tau)-G^{-}(x,y,\tau)\right]\label{EQ:GreenK_gammainf}
\end{equation}

\begin{equation}
G^{\pm}(x,y,\tau)=\left(\frac{\sin\frac{\pi(x+y+iv\tau)}{2L}\sin\frac{\pi(x+y-iv\tau)}{2L}}{\sin\frac{\pi(x-y+iv\tau)}{2L}\sin\frac{\pi(x-y-iv\tau)}{2L}}\right)^{\pm K}\phi_{\pm}(x,y).\label{EQ:GreenPMK_gammainf}
\end{equation}

\[
\phi_{\pm}(x,y)=\cos\left(\frac{\pi(x\mp y)}{L}\right)\text{ for odd, 1 for even.}
\]

Correspondingly, the dimensionless versions of Eqs.~(\ref{EQ:GreenK_gammainf},\ref{EQ:GreenPMK_gammainf})
are:

\begin{align}\label{EQ:GreenK_gammainfR}
 & G(\widetilde{x},\widetilde{y},\widetilde{\tau})=\frac{A_{z}}{2}\left(\frac{2L}{\pi}\right)^{-2K}\left(\sin{\pi\widetilde{x}}\sin{\pi\widetilde{y}}\right)^{-K}\left[\mathcal{G}^{+}(\widetilde{x},\widetilde{y},\widetilde{\tau})-\mathcal{G}^{-}(\widetilde{x},\widetilde{y},\widetilde{\tau})\right]\nonumber \\
 & \mathcal{G}^{\pm}(x,y,\tau)=\left(\frac{\sin\frac{\pi(\tilde{x}+\tilde{y}+i\gamma\tilde{\tau})}{2}\sin\frac{\pi(\tilde{x}+\tilde{y}-i\gamma\tilde{\tau})}{2}}{\sin\frac{\pi(\tilde{x}-\tilde{y}+i\gamma\tilde{\tau})}{2}\sin\frac{\pi(\tilde{x}-\tilde{y}-i\gamma\tilde{\tau})}{2}}\right)^{\pm K}\varphi_{\pm}(\tilde{x},\tilde{y})\\
 & \varphi_{\pm}(\tilde{x},\tilde{y})=\cos\pi(\tilde{x}\mp\tilde{y})\text{ for odd, 1 for even.}\nonumber
\end{align}

Here comes our \textbf{assumption 2}, based on the limiting properties
of $\theta-$ functions and the known isotropic case solution. \textit{We
propose the following generalization of terms in~(\ref{EQ:GreenK_gammainfR})
for arbitrary $\gamma$}:

\begin{align}\label{EQ:FullGreen_T23}
 & \cos\pi(\tilde{x}\mp\tilde{y})\text{ for odd, 1 for even }\Rrightarrow\frac{\theta_{2,3}(\pi(\tilde{x}\mp\tilde{y}),e^{-\frac{\pi\gamma}{2K}})}{\theta_{2,3}(0,e^{-\frac{\pi\gamma}{2K}})},\nonumber \\
 & \left(\frac{\sin\frac{\pi(\tilde{x}+\tilde{y}+i\gamma\tilde{\tau})}{2}\sin\frac{\pi(\tilde{x}+\tilde{y}-i\gamma\tilde{\tau})}{2}}{\sin\frac{\pi(\tilde{x}-\tilde{y}+i\gamma\tilde{\tau})}{2}\sin\frac{\pi(\tilde{x}-\tilde{y}-i\gamma\tilde{\tau})}{2}}\right)^{\pm K}\Rrightarrow\left(\frac{\theta_{1}(\frac{\pi(\tilde{x}+\tilde{y}+i\gamma\tilde{\tau})}{2},e^{-\frac{\pi\gamma}{2}})\theta_{1}(\frac{\pi(\tilde{x}+\tilde{y}-i\gamma\tilde{\tau})}{2},e^{-\frac{\pi\gamma}{2}})}{\theta_{1}(\frac{\pi(\tilde{x}-\tilde{y}+i\gamma\tilde{\tau})}{2},e^{-\frac{\pi\gamma}{2}})\theta_{1}(\frac{\pi(\tilde{x}-\tilde{y}-i\gamma\tilde{\tau})}{2},e^{-\frac{\pi\gamma}{2}})}\right)^{\pm K}\\
 & \left(\sin{\pi\widetilde{x}}\sin{\pi\widetilde{y}}\right)^{-K}\Rrightarrow\left(\frac{\theta_{1}(\pi\widetilde{x},e^{-\frac{\pi\gamma}{2}})\theta_{1}(\pi\widetilde{y},e^{-\frac{\pi\gamma}{2}})}{[\partial_{x}\theta_{1}(0,e^{-\frac{\pi\gamma}{2}})]^{2}}\right)^{-K}\nonumber
\end{align}
\end{widetext}

The Green function in question was not derived rigorously, so (\ref{EQ:FullGreen_T23})
should be considered as some sort of \emph{ansatz}. For $K=1/2$ case
this recovers Eqs.~(\ref{EQ:EAH_Green_reduced}, \ref{EQ:SQRTthingPtilde},\ref{EQ:SQRTthingMtilde})
verbatim. The left-hand side is the correct $\gamma\rightarrow\infty$
limit of the right-hand side.

\subsubsection{Universal scaling}

Hence, from our assumptions 1 and 2 we can conclude that the staggered
transverse susceptibility per TLL site has the following universal form
(with all the SI factors now recovered):
\begin{widetext}
\begin{equation}
\chi_{\pi}^{\pm}(L,T)=(g\mu_{B})^{2}\frac{A_{x}}{k_{B}v}\left(\frac{T}{v}\right)^{1/2K-2}F_{K}^{e,o}\left(\frac{v}{LT}\right),\label{EQ:FiniteSizeSusc}
\end{equation}

where

\begin{equation}
F_{K}^{e,o}(\gamma)=\frac{1}{2}\gamma^{1/2K-1}\left(\frac{\pi}{2}\right)^{1/2K}\iiint\limits _{0}^{1}\mathcal{G}_{K,\gamma}(\widetilde{x},\widetilde{y},\widetilde{\tau})d\widetilde{x}d\widetilde{y}d\widetilde{\tau},\label{EQ:chi_universcaling}
\end{equation}

\begin{align}\label{EQ:MegafinalG}
\mathcal{G}_{K,\gamma}(\widetilde{x},\widetilde{y},\widetilde{\tau})=\left(\frac{\theta_{1}(\pi\widetilde{x},e^{-\frac{\pi\gamma}{2}})\theta_{1}(\pi\widetilde{y},e^{-\frac{\pi\gamma}{2}})}{[\partial_{x}\theta_{1}(0,e^{-\frac{\pi\gamma}{2}})]^{2}}\right)^{-1/4K}\times\phantom{-----}\nonumber \\
\Bigg\{\frac{\theta_{2,3}(\pi(\tilde{x}-\tilde{y}),e^{-2K\pi\gamma})}{\theta_{2,3}(0,e^{-2K\pi\gamma})}\left(\frac{\theta_{1}(\frac{\pi(\tilde{x}+\tilde{y}+i\gamma\tilde{\tau})}{2},e^{-\frac{\pi\gamma}{2}})\theta_{1}(\frac{\pi(\tilde{x}+\tilde{y}-i\gamma\tilde{\tau})}{2},e^{-\frac{\pi\gamma}{2}})}{\theta_{1}(\frac{\pi(\tilde{x}-\tilde{y}+i\gamma\tilde{\tau})}{2},e^{-\frac{\pi\gamma}{2}})\theta_{1}(\frac{\pi(\tilde{x}-\tilde{y}-i\gamma\tilde{\tau})}{2},e^{-\frac{\pi\gamma}{2}})}\right)^{1/4K}-\\
\frac{\theta_{2,3}(\pi(\tilde{x}+\tilde{y}),e^{-2K\pi\gamma})}{\theta_{2,3}(0,e^{-2K\pi\gamma})}\left(\frac{\theta_{1}(\frac{\pi(\tilde{x}+\tilde{y}+i\gamma\tilde{\tau})}{2},e^{-\frac{\pi\gamma}{2}})\theta_{1}(\frac{\pi(\tilde{x}+\tilde{y}-i\gamma\tilde{\tau})}{2},e^{-\frac{\pi\gamma}{2}})}{\theta_{1}(\frac{\pi(\tilde{x}-\tilde{y}+i\gamma\tilde{\tau})}{2},e^{-\frac{\pi\gamma}{2}})\theta_{1}(\frac{\pi(\tilde{x}-\tilde{y}-i\gamma\tilde{\tau})}{2},e^{-\frac{\pi\gamma}{2}})}\right)^{-1/4K} & \Bigg\}.\nonumber
\end{align}

The power law $T^{\frac{1}{2K}-2}$ is the key result. It also matches
the infinite length limit power law for the staggered susceptibility~\cite{Giamarchi2003,Bouillot2011}:

\begin{equation}
\chi_{\pi}^{\pm}(T)=(g\mu_{B})^{2}\frac{A_{x}}{k_{B}v}\left(\frac{T}{v}\right)^{1/2K-2}\times\left[\frac{1}{2}(2\pi)^{1/2K-2}\sin\left(\frac{\pi}{4K}\right)B^{2}\left(\frac{1}{8K},1-\frac{1}{4K}\right)\right].\label{EQ:Thermlimit}
\end{equation}
\end{widetext}

\section{Universal susceptibility vs numerics and experiment}

In this section we will compare the previously developed analytical
ansatz with the results of QMC calculations.

\subsection{The QMC data and the pristine TLL description}

\begin{figure*}
\centering \includegraphics[width=1\textwidth]{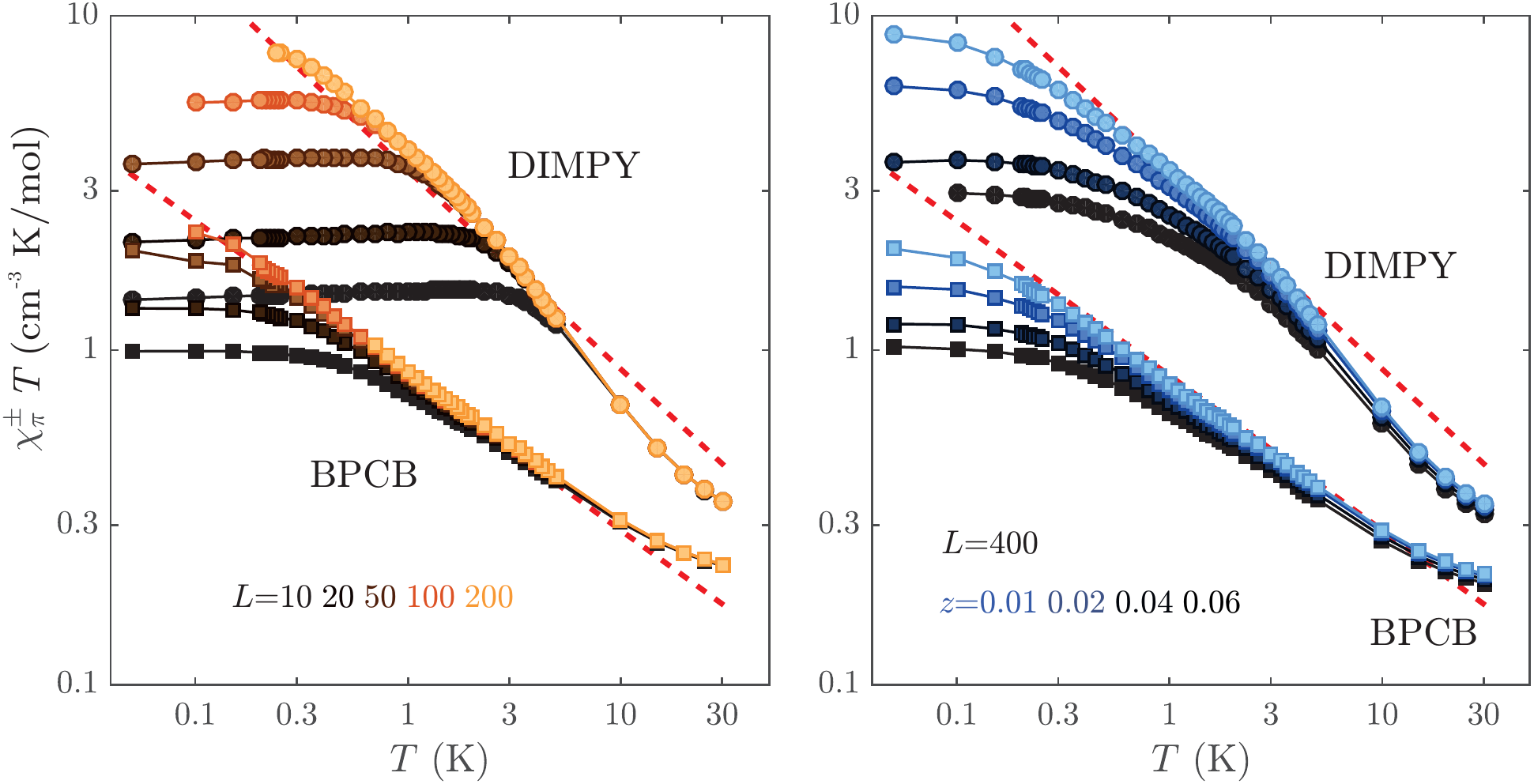} \caption{The QMC susceptibility data for DIMPY and BPCB. In the left panel
the data for the finite-size ladder segments is shown; in the right
panel --- for the long segments with depletion $z$. Dashed lines
show the TLL result for the clean infinite system..}
\label{fig:QMCbare}
\end{figure*}

In Fig.~\ref{fig:QMCbare} the staggered susceptibility versus temperature
is displayed for all the types of ladders considered: both BPCB and
DIMPY at $8$~T, in shorter finite-segment ($10-200$ rungs) and
depleted long segment ($400$ rungs) settings.
This data can be compared to the expectations for the corresponding
pristine TLL model. One can see that in all the cases the ``paramagnetic''
regime at high temperatures is followed by the part that matches the
$L=\infty$ TLL behavior~(\ref{EQ:Thermlimit}) well. This temperature
window gets shorter as the segment size is decreased or more disorder
is added, and at low $T$ the ``finite-size'' behavior starts to
dominate the susceptibility.

\subsection{Scaled QMC data and the scaling ansatz}

\begin{figure*}
\centering \includegraphics[width=1\textwidth]{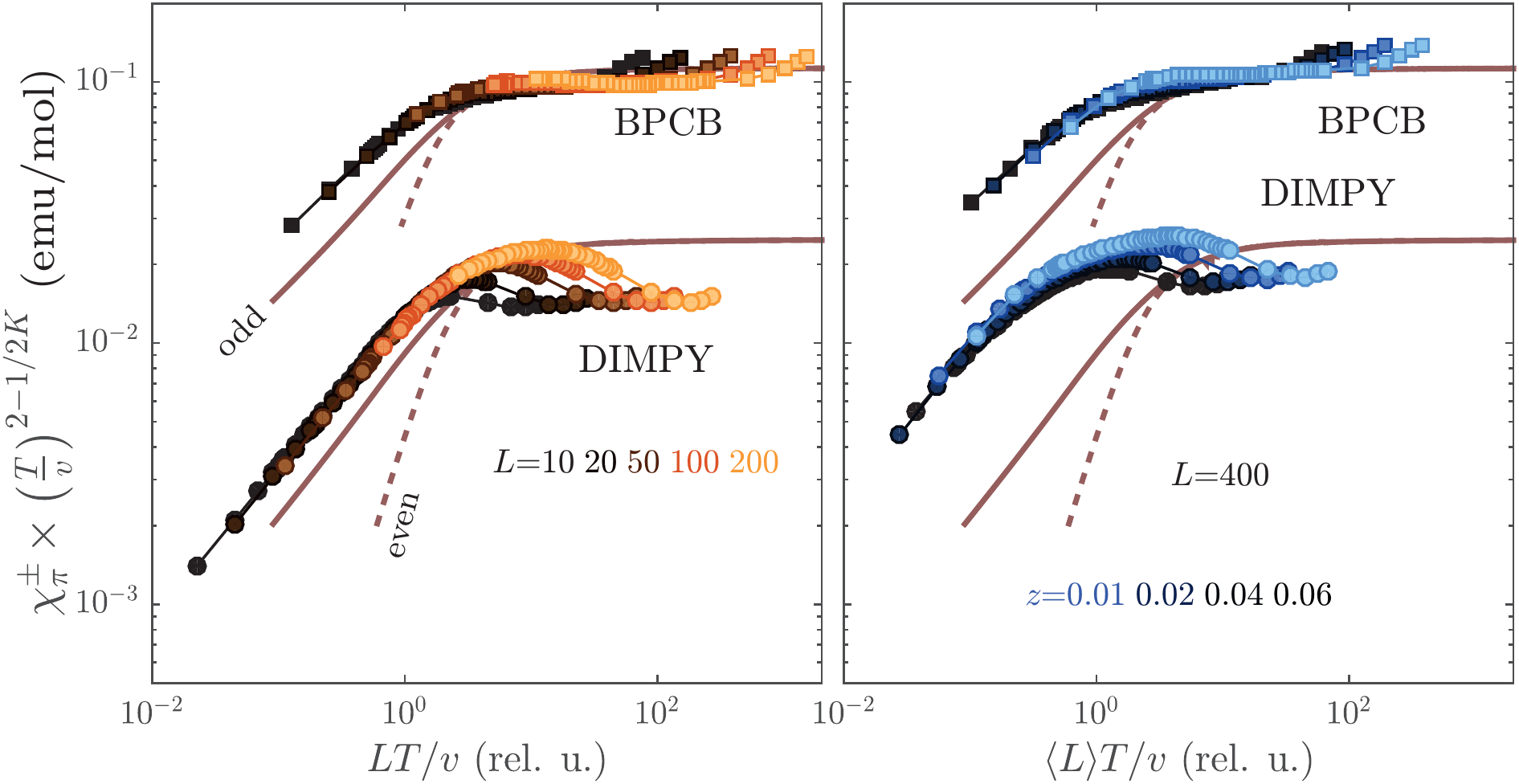} \caption{Scaled QMC data for DIMPY and BPCB versus Eqs.~(\ref{EQ:FiniteSizeSusc}-\ref{EQ:MegafinalG}).
In the left panel the data for the finite-size ladder segments is
shown; in the right panel --- for the long segments with depletion
$z$. The analytical curves are identical in both cases. Solid line
stands for the odd version of Eqs.~(\ref{EQ:FiniteSizeSusc}-\ref{EQ:MegafinalG}),
and dashed line for the even one.}
\label{fig:QMCscaled}
\end{figure*}

The results can further be compared to the predictions of $LT/v$
scaling behavior. The key result of our ansatz is contained in Eqs.~(\ref{EQ:FiniteSizeSusc}-\ref{EQ:MegafinalG}).
These functions for both BPCB and DIMPY cases are plotted along with
the data for the finite-size segments and for the long depleted segments.
We can see in Fig.~\ref{fig:QMCscaled} that in general the behavior
of the scaled staggered susceptibility is in the quantitative agreement
with the theoretical predictions. However, there is a number of discrepancies.

First, only the ``odd'' version of the scaling function $F_{K}(\gamma)$~(\ref{EQ:chi_universcaling})
seems to be relevant to the data. This is understood as a consequence
of extracting the $\chi_{\pi}^{\pm}(T)$ in an indirect way from the
equal-time correlations. The latter quantity does not display an even-odd
sensitivity, so neither does the obtained susceptibility.

Second, in the shorter finite-sized segments we see an offset in $LT/v$,
by the factor of $2$ roughly. This offset seems to be the same for
both DIMPY and BPCB datasets. The reason for that is not understood
now.

\begin{figure*}
\centering \includegraphics[width=1\textwidth]{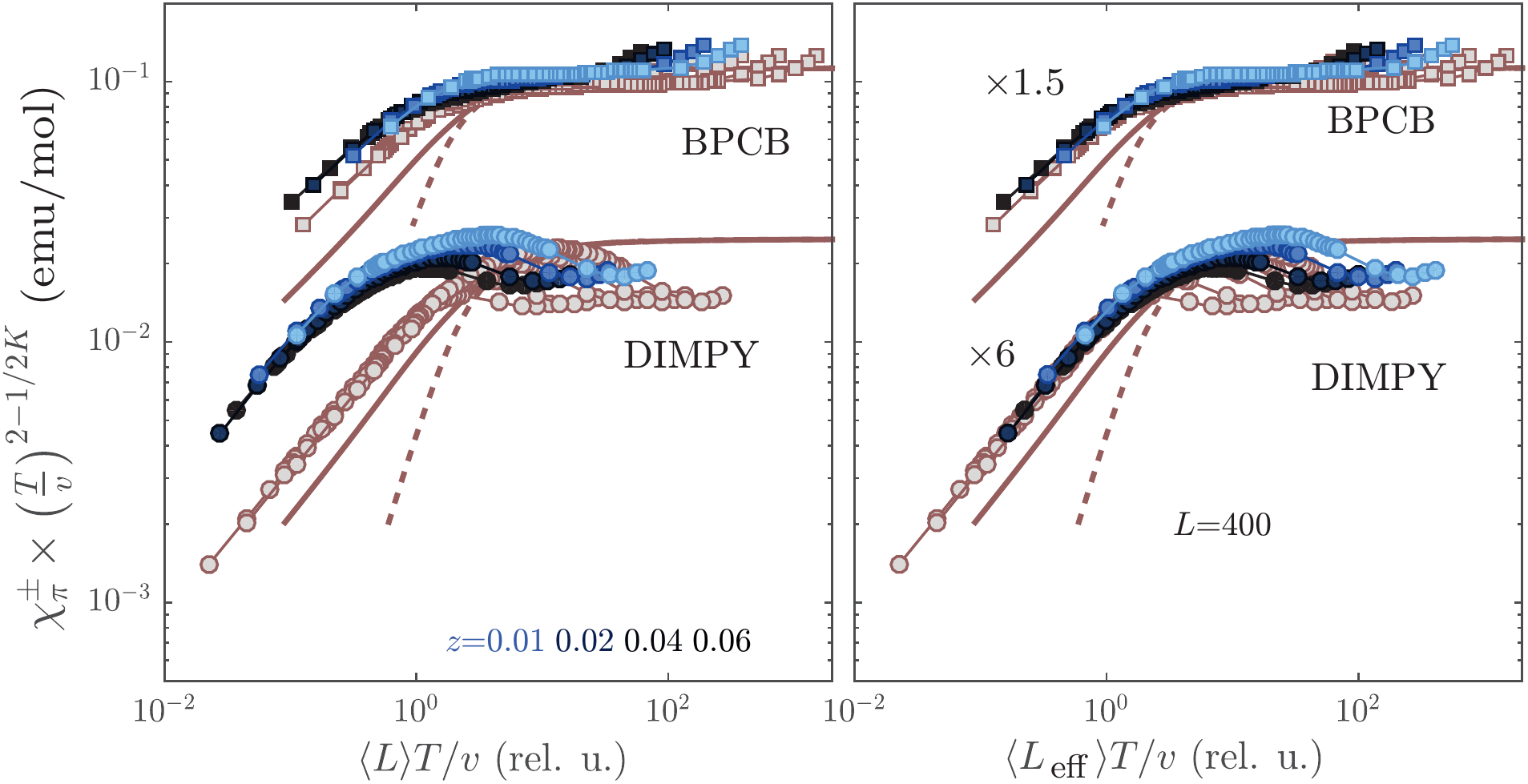} \caption{Comparison of the scaled datasets for the depleted ladders and finite-sized
segments. The latter are shown in the background together with the
analytical ansatz curves. In the left panel the data is shown ``as
is'', and in the right panel the average impurity-impurity distance
is enlarged by the effective factor $\lambda_{\text{eff}}$ ($1.5$
for BPCB, $6$ for DIMPY).}
\label{fig:QMCscaledshifted}
\end{figure*}

Third, we see a much more significant offset in the depleted ladders
datasets. This time the offset \emph{does} depend on the particular
ladder (DIMPY or BPCB), being much stronger for the strong-leg case,
as the Fig.~\ref{fig:QMCscaledshifted} shows. We understand it as
the effect of the partial defect transparency discussed in the main
text. This partial transparency means that the effective length has
to be renormalized: it is not merely a mean distance between the defects
but some longer effective scale. Naturally, this effect is much stronger for
the case of DIMPY (where the defects are very transparent), while
in BPCB this renormalization is insignificant compared to the bare
segment case. As Fig.~\ref{fig:QMCscaledshifted} shows, for BPCB
the length renormalization factor is mere $1.5$, while it is $6$
for DIMPY.

\subsection{The scaling ansatz and the measured phase diagrams}

Since the scaling predictions are in a good agreement with the predictions
of the numerical simulations, one can try to predict the depleted
ladder phase diagram in ``bruteforce'' manner, based only on the
known field dependencies of $K$, $v$ and $A_{x}$. The susceptibility
of an infinite depleted ladder can be obtained from taking the distribution
of the segment length into account for a given $z$. An empirical
correction factor for the effective length dicussed in the previous
section can also be included. Then, susceptibility of an individual
segment of a given length can be obtained with the Eqs.~(\ref{EQ:FiniteSizeSusc}-\ref{EQ:MegafinalG}). Thus, the susceptibility of the depleted ladder per mol of spins is:

\begin{equation}
\chi_{\pi}^{\pm}(z,T)=N_{A}\sum_{L}\rho_{z}(L)\chi_{\pi}^{\pm}(L\lambda_{\text{eff}},T).\label{EQ:PoissonSum}
\end{equation}

with $\lambda_{\text{eff}}$ being the parameter, describing the effective
length of the segment between the two impurities at distance $L$
from each other, The probability of finding such segment is given
by:

\begin{equation}
\rho_{z}(L)=L(2z)^{2}(1-2z)^{L}.\label{EQ:lengthprobab}
\end{equation}

\begin{figure}
\centering \includegraphics[width=0.5\textwidth]{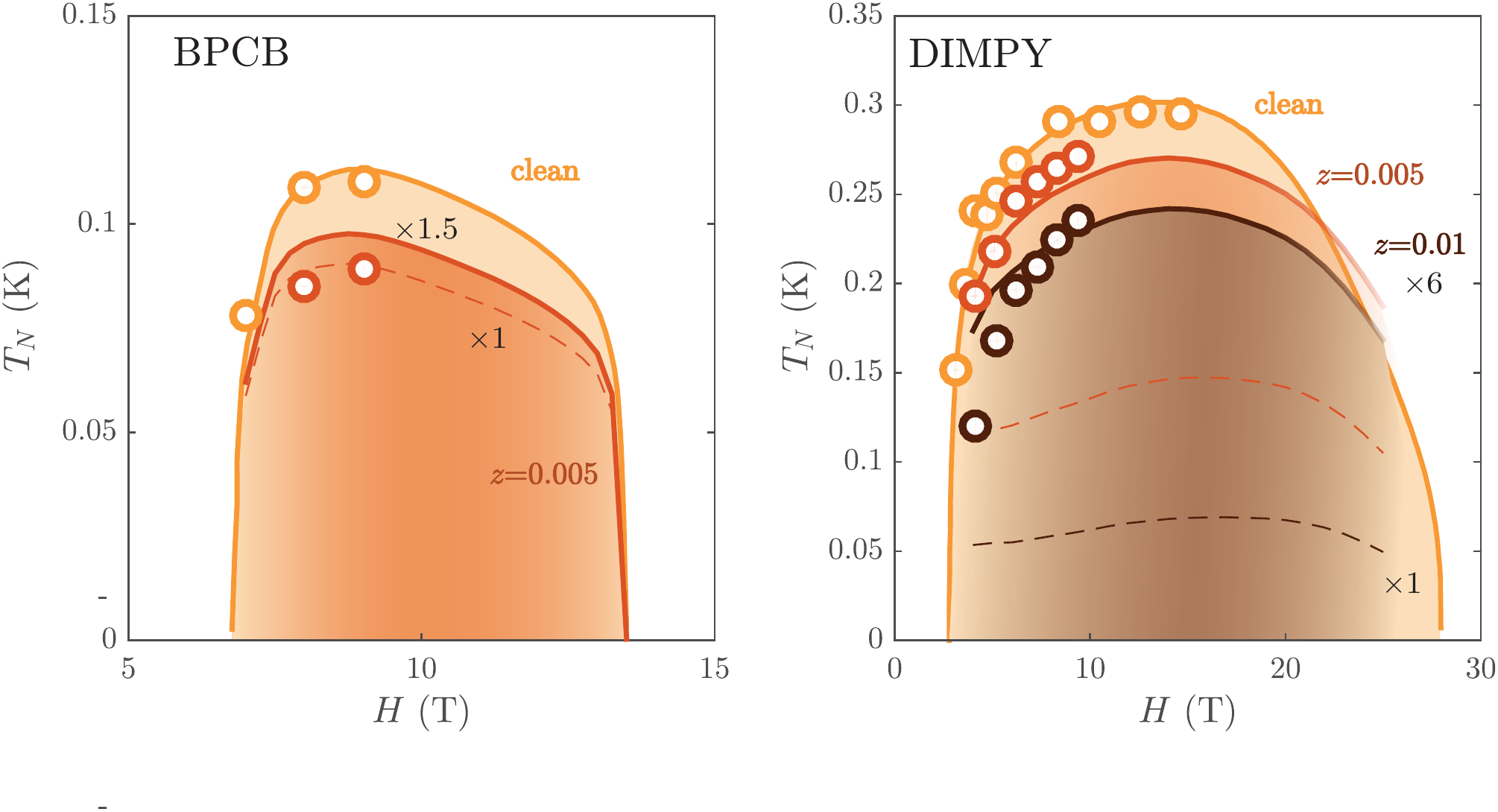} \caption{Experimental data vs TLL scaling-based approach for DIMPY and BPCB.
The color coding is the same as in Fig.~5 from the main text; the
experimental data points are also identical. Sold lines stand for
the theoretical calculation~(\ref{EQ:PoissonSum}) taking the $\lambda_{\text{eff}}$
into account (as also indicated on the plot). Dashed lines show the
result that does not include this renormalization correction. The
``clean'' lines is the infinite system TLL result, known from the
literature. }
\label{fig:ScalingPhD}
\end{figure}

Using the known $T_{N}(H)$ in the pristine TLL model and the field dependencies of effective TLL parameters as given Refs.~\cite{Bouillot2011,Schmidiger2012},
we obtain the approximations of the phase boundaries shown in Fig.~\ref{fig:ScalingPhD}.

We see that overall the agreement between the analytical approximation
and the observed phase boundaries is quite good. The $\lambda_{\text{eff}}$
correction is rather marginal in case of BPCB. In case of DIMPY, however,
it is absolutely crucial for obtaining the realistic result. We conclude
that the approach based on the $LT/v$ universality and the mean field
ordering criterion works quite well for describing the phase diagrams
of the depleted spin ladders.

\bibliography{Bibliography_modKirill}